\documentclass[pdflatex,sn-mathphys-ay]{sn-jnl}

\usepackage{tcolorbox}

\usepackage{amsmath,amssymb,amsfonts}
\usepackage{graphicx}

\usepackage{booktabs}
\usepackage{multirow}
\usepackage{bm}
\usepackage{mathtools}
\usepackage{url}
\usepackage{xcolor}

\usepackage{amsfonts}
\usepackage{amsmath}

\usepackage{natbib}

\usepackage{algorithm}
\usepackage{algpseudocode}
\usepackage{float}
\algrenewcommand\algorithmicrequire{\textbf{Input:}}
\algrenewcommand\algorithmicensure{\textbf{Output:}}

\let\oldmarginpar\marginpar
\renewcommand{\marginpar}[1]{\oldmarginpar{\scriptsize #1}}

\usepackage[title]{appendix}%

\title[Robust estimation under heavy tails]{Calibration of robust covariance estimation under heavy tails}

\author*[1]{\fnm{Andrea} \sur{Cerioli}}\email{andrea.cerioli@unipr.it}

\author[2]{\fnm{Domenico} \sur{Perrotta}}\email{domenico.perrotta@ec.europa.eu}

\author[2]{\fnm{Francesca} \sur{Torti}}\email{francesca.torti@ec.europa.eu}

\affil*[1]{\orgdiv{Department of Economics and Management}, \orgname{University of Parma}, \orgaddress{\street{via Kennedy 6}, \city{Parma}, \postcode{43125}, 
\country{Italy}}}

\affil[2]{\orgdiv{Joint Research Centre (JRC)}, \orgname{European Commission}, \orgaddress{\street{via Fermi 2749}, \city{Ispra (VA)}, \postcode{21027}, 
\country{Italy}}}

\date{}

\begin{document}













\abstract{The Minimum Covariance Determinant (MCD) estimator is one of the most widely used high-breakdown estimators of multivariate location and scatter. Its practical success relies on suitable corrections, which remove the asymptotic and small-sample biases introduced by trimming. While explicit consistency factors are available under the multivariate normal distribution and, more recently, under multivariate Student-\textit{t} models, these corrections only address the asymptotic behavior of the estimator.
In the Gaussian case, it is known showed that the MCD also exhibits a non-negligible finite-sample bias, particularly for small sample sizes and relatively large dimensions. Such bias propagates to robust Mahalanobis distances and may affect subsequent procedures, including multivariate outlier detection and robust inference.
This paper investigates the finite-sample behavior of the MCD estimator and of the related robust distances under multivariate Student-\textit{t} distributions, a potentially relevant scenario in many application fields. Extensive Monte Carlo experiments are carried out over a broad range of sample sizes, dimensions, trimming levels and degrees of freedom. Based on these experiments, empirical correction factors are obtained and approximated by smooth functions, providing a computationally efficient calibration suitable for practical implementations to heavy-tailed data. The proposed corrections are developed both for the scatter bias and the robust distance quantiles, thus leading to improved approximations for robust inference. 
}




\maketitle

\section{Introduction}

Robust estimation of multivariate location and scatter is a fundamental
problem in multivariate statistics. Reliable estimates of these
quantities are required in a wide range of statistical procedures,
including principal component analysis, discriminant analysis,
multivariate regression, clustering and multivariate outlier detection.
When a portion of the data depart from the postulated multivariate model, most often the normal one, classical estimators based on the sample mean and covariance matrix may become unreliable. This has motivated the
development of robust alternatives with improved breakdown properties
\citep{hub+al:08,hub+al:18,mar+al:19}.

Among the robust procedures proposed in the literature, the Minimum
Covariance Determinant (MCD) estimator occupies a prominent position.
Introduced by \cite{rou:84} and further studied by
\cite{but+al:93}, \cite{cr+h:99}, \cite{rous+d:99} and \cite{cat+lop:12}, among others, the MCD combines a high breakdown point, affine equivariance and computational feasibility. These properties have made it one of the reference estimators in robust multivariate analysis, with applications ranging from robust multivariate analysis to outlier detection based on robust Mahalanobis distances, also in high dimensions \citep{cer:10,cer+far:11, rou+hub:11,far+gre:15,cla+gro:23,pfe+al:25,ley+al:26,nor+al:26,puc+al:26}.

The MCD estimator is obtained by selecting a subset of observations whose
empirical covariance matrix has minimum determinant. Because trimming
systematically modifies the scale of the retained observations, a
consistency correction is required before the estimator can be used for
inference or for the computation of robust distances. Under multivariate
normality of the generating process of the uncontaminated observations, asymptotic consistency factors have long been available, while
\cite{pis+al:02} introduced further finite-sample correction factors that improve the calibration of the estimator for small and moderate sample sizes. 
These corrections have become standard in practical implementations of the MCD: popular ones in \texttt{R}, \texttt{Python} and \texttt{MATLAB} are by \citet{tod+fil:09}, \citet{RobPy:2024}, \citet{fsda:2012,fsda:2015} and \citet{FSDA:toolbox}.

Although the Gaussian model remains a central reference in robust
statistics, many applications involve distributions with heavier tails for the generating process of the uncontaminated observations. Elliptically contoured distributions provide a natural candidate for extending the multivariate normal model while preserving its geometric structure \citep{fan+al:90,ars:04,gup+al:13,fil+kol:24,per+al:24}. In particular, the multivariate Student-\textit{t} distribution represents a flexible model for data characterized by moderate or pronounced tail heaviness.
A few recent contributions have thus extended the theoretical calibration of the MCD beyond the Gaussian setting: \cite{bar+al:23} derived the asymptotic consistency factor of the MCD under multivariate Student-\textit{t} distributions, 
while \cite{bar+al:26} developed robust distances and multivariate outlier detection procedures that explicitly account for heavy-tailed populations.

The availability of the Student-\textit{t} consistency factor resolves the asymptotic calibration problem, but it does not automatically guarantee an accurate finite-sample behavior. In fact, the Gaussian results of \cite{pis+al:02} show that, after consistency correction, residual
finite-sample effects may still affect the scatter estimate. Since robust distances and related inferential procedures depend directly on the estimated scatter matrix, these effects may propagate to subsequent analyses and inferences, as shown by \cite{cra:09}. A corresponding \textit{finite-sample calibration for heavy-tailed elliptical models} is currently unavailable. 

This calibration is the target of the present work, which spans two complementary directions. One direction specifically covers the scatter matrix bias, while the second one addresses the tail distribution of the squared robust distances that are the main inferential tool for outlier detection. In both cases, through extensive simulation and careful interpolation, we provide results that cover most of the situations of practical interest where calibration is actually needed. We also make our routines available and our results fully reproducible. In the same spirit, our routines provide the tools to extend the calibration study outside the specific grid that we have chosen. Therefore, our approach can be extended to virtually any sample size, dimension, and trimming level of potential interest to the user.

Throughout the paper, we use the term ``finite-sample calibration'' rather than ``small-sample correction'', since the proposed adjustment is not restricted to very small sample sizes. Our aim is to improve the calibration of the estimator for finite values of the sample size, bridging the gap between asymptotic consistency and practical implementation. Furthermore, our calibration principle is general and could be extended to other members of the class of affine-equivariant trimmed estimators, in which the MCD is a prominent example. 

%

The remainder of the paper is organized as follows. 
Section \ref{Preliminaries:sec} 
recalls the framework of trimmed estimators and the asymptotic Student-\textit{t} consistency correction. 
Section \ref{calibration:sec} develops the finite-sample calibration principle and its determinant-based implementation for the MCD. 
Section \ref{MC:sec} describes the Monte Carlo study used to estimate the correction factors. Section \ref{finite-sample:sec} details the corresponding algorithm and presents the results from different points of view. Section \ref{sec:distance_calibration} reports the methodology and results for the calibration of the squared robust distances. 
Section \ref{data:sec} shows the usefulness of the proposed corrections by implementing them in a real data example. Finally, Section \ref{sec:simulation_implementation} illustrates both the implementation of our algorithms and how our results can be reproduced.
The Appendices serve the purpose to complement several aspects of the methodology considered in the text.
%
%



\section{Preliminaries}
\label{Preliminaries:sec}

Affine-equivariant 
trimmed estimators constitute a broad class of robust procedures for multivariate location and scatter estimation.
Following the framework adopted by \citet{bar+al:26}, let $\mathbf{X}_1,\ldots,\mathbf{X}_n$ be a random sample from a $p$-variate elliptical distribution with location vector $\boldsymbol{\mu}$ and population covariance matrix $\boldsymbol{\Sigma}$.
Let $\alpha_n\in(0,0.5)$ denote the trimming proportion used for a sample of size $n$. Associated with $\alpha_n$, let $w_{i,\alpha_n}\in\{0,1\}$, $i=1,\ldots,n$, denote binary trimming weights, where
$w_{i,\alpha_n}=1$ if observation $\mathbf{X}_i$ is retained and $w_{i,\alpha_n}=0$ otherwise. Define
\begin{equation}\label{wn}
w_{\alpha_n}=\sum_{i=1}^{n}w_{i,\alpha_n},    
\end{equation}
to be the number of retained observations.
The corresponding 
trimmed estimators of location and scatter are
\begin{equation}
\widetilde{\boldsymbol{\mu}}_{\alpha_n}
=
\frac{1}{w_{\alpha_n}}
\sum_{i=1}^{n}
w_{i,\alpha_n}\mathbf{X}_i,
\label{eq:trimmean}
\end{equation}
and
\begin{equation}
\widetilde{\boldsymbol{\Sigma}}_{\alpha_n}
=
\frac{\kappa}{w_{\alpha_n}}
\sum_{i=1}^{n}
w_{i,\alpha_n}
\left(
\mathbf{X}_i-
\widetilde{\boldsymbol{\mu}}_{\alpha_n}
\right)
\left(
\mathbf{X}_i-
\widetilde{\boldsymbol{\mu}}_{\alpha_n}
\right)^{\!\top},
\label{eq:trimcov}
\end{equation}
where $\kappa$ denotes a consistency factor ensuring that
$\widetilde{\boldsymbol{\Sigma}}_{\alpha_n}$ estimates the target population covariance matrix when the observations are generated from the assumed model.
Different choices of the trimming weights lead to different robust estimators within this general framework.
Among these methods, the MCD estimator is the one considered in the present work. 

The MCD is 
obtained by selecting the subset of observations whose empirical covariance matrix has minimum determinant. If the selected subset contains $h_n$ observations, then the MCD weights satisfy
\[
h_n=\sum_{i=1}^{n}w_{i,\alpha_n}=w_{\alpha_n}.
\]
Thus, in the MCD case, the general notation $w_{\alpha_n}$ introduced in \eqref{wn} corresponds to the subset size $h_n$.
The actual level of trimming in the MCD can also be specified through the
asymptotic trimming proportion
\[
\alpha_0=\lim_{n\rightarrow\infty}\alpha_n .
\]
For a chosen value of $\alpha_0 \in(0,0.5]$, the finite-sample subset size is obtained
through the standard adjustment formula \citep{mac:26}
\begin{equation}
\label{eq:hn}
h_n=
\left\lfloor
n-2(n-q_n)\alpha_0 \right\rfloor,
\end{equation}
where
\[
q_n=
\left\lfloor\frac{n+p+1}{2}\right\rfloor .    
\]
Consequently,
\begin{equation}
\label{eq:alphan}
\alpha_n=1-\frac{h_n}{n}.    
\end{equation}
Although not explicitly included in our notation, the dependence of $h_n$ on $p$ reflects the fact that, in the MCD, the definition of the finite-sample trimming level involves the dimension of the problem.

If $\mathbf{X}_1,\ldots,\mathbf{X}_n$ follow a $p$-variate normal law with mean vector $\mathbf{\mu}$ and covariance matrix $\mathbf{\Sigma}$, the consistency factor in Equation \eqref{eq:trimcov} is the well known expression
\begin{equation}\label{eq:consistN}
\kappa \equiv \eta_{\alpha_0,p}
=
\frac{1-\alpha_0}
{F_{\chi^2_{p+2}}
\left(
F^{-1}_{\chi^2_p}(1-\alpha_0)
\right)},
\end{equation}
where $F_{\chi^2_q}$ denotes the distribution function of a chi-square
random variable with $q$ degrees of freedom. 

\citet{bar+al:23} extended the above result to the multivariate Student-\textit{t} law with location parameter $\boldsymbol{\mu}$, scale matrix $\boldsymbol{\Psi}$ and $\nu$ degrees of freedom, here denoted as $t_p(\boldsymbol{\mu},\boldsymbol{\Psi},\nu)$. If 
$\mathbf{X}_1,\ldots,\mathbf{X}_n$ are now independent replicates of the random vector
\begin{equation}\label{eq:assT}
\mathbf{X}
\sim
t_p(\boldsymbol{\mu},
\boldsymbol{\Psi},
\nu) ,
\end{equation}
and if $\nu>2$, so that the covariance matrix of $\mathbf{X}$ is
\begin{equation}\label{eq:covT}
\mathbf{\Sigma}
=
\frac{\nu}{\nu-2}\mathbf{\Psi} ,
\end{equation}
the scaling factor making
$\widetilde{\boldsymbol{\Sigma}}_{\alpha_n}$ a consistent estimator of
$\boldsymbol{\Sigma}$ 
becomes 
\begin{equation}\label{eq:consistT}
\kappa \equiv \eta_{\alpha_0,p}(\nu)
=
\left[
\frac{\nu-2}{(1-\alpha_0)p}
\int_0^{1-\alpha_0}
\frac{du}
{1-I^{-1}_{u}(p/2,\nu/2)}
-
\frac{\nu-2}{p}
\right]^{-1} .
\end{equation}
Here $I^{-1}_{u}(a,b)$ is the inverse of the regularized incomplete beta function and our notation stresses the dependence of 
$\kappa$ 
also on the tail parameter $\nu$ of the underlying distribution. Furthermore,
$\eta_{\alpha_0,p}(\nu)$ in \eqref{eq:consistT} reduces to
$\eta_{\alpha_0,p}$ in \eqref{eq:consistN} as $\nu\rightarrow\infty$.
%

The availability of $\eta_{\alpha_0,p}(\nu)$ provides the asymptotic calibration of the MCD estimator under a multivariate Student-\textit{t} model. Since the consistency factor is independent of the sample size, however, its ability to provide an adequate calibration for moderate sample sizes remains an open question, which is investigated in the following sections.
%

\section{Finite-sample calibration}
\label{calibration:sec}



Rather than viewing the calibration problem as specific to the MCD, it is useful to
consider it within the broader class of 
trimmed estimators
introduced in Section~2.
Let
$\widetilde{\boldsymbol{\Sigma}}_{\alpha_n}$
denote an asymptotically calibrated estimator of scatter, namely an
estimator already corrected by the appropriate consistency
factor for a given asymptotic trimming level $\alpha_0$, as in Equation \eqref{eq:trimcov}.
For finite samples, one may seek an additional multiplicative
calibration factor,
denoted by
$\delta_n$,
and consider the estimator
\begin{equation}
\widehat{\boldsymbol{\Sigma}}_{\alpha_n}
=
\delta_n
\widetilde{\boldsymbol{\Sigma}}_{\alpha_n},
\label{eq:deltacorrection}
\end{equation}
with the constraint that
\(
\delta_n \rightarrow 1
\)
as $n\rightarrow \infty$. Therefore, the role of $\delta_n$ is not to modify the asymptotic behavior of the estimator but rather to reduce the residual finite-sample discrepancy that remains after the consistency correction has been applied.
Conceptually, the calibration factor may be defined with respect to a loss function measuring the distance between the corrected estimator and the target population covariance matrix.
Different choices of the loss lead to different calibration criteria, depending on the inferential objective and on the estimator under consideration, the MCD in the present work.

Since the MCD is defined through the minimization of the determinant of the empirical scatter matrix, the determinant provides a natural diagnostic quantity for studying the finite-sample behavior of this estimator \citep{pis+al:02}. Accordingly, the proposed calibration strategy consists of determining, by Monte Carlo experiments, the value of \(\delta_n\) that compensates for the residual finite-sample bias after the asymptotic consistency correction has been applied. 
Under the multivariate Student-\textit{t} model \eqref{eq:assT}, a natural determinant-based measure of the residual finite-sample
bias is
\begin{equation}\label{eq:BiasDet}
    E\{\widetilde{d}_{p,n,\nu}^{\alpha_0}\} ,
\end{equation}
where
\begin{equation}\label{eq:detdia}
\widetilde{d}_{p,n,\nu}^{\alpha_0}
=
\left(
\left|
\boldsymbol{\Sigma}^{-1}
\widetilde{\boldsymbol{\Sigma}}_{\alpha_n}
\right|
\right)^{1/p}
\end{equation}
is the determinant-based 
diagnostic computed from the MCD scatter estimator $\widetilde{\boldsymbol{\Sigma}}_{\alpha_n}$ after application of the asymptotic Student-\(t\) consistency factor \(\eta_{\alpha_0,p}(\nu)\), and $\boldsymbol{\Sigma}$ is the target matrix \eqref{eq:covT}. As in \cite{pis+al:02}, we are interested in correcting the bias in the scale of the MCD estimator of scatter and leave the investigation on its shape to future research \citep[see, e.g.,][]{pai+van:14}.

Some simple algebra (see Appendix \ref{recall_detdia:sec}) yields
\begin{equation}\label{eq:BiasDet_full}
\widetilde{d}_{p,n,\nu}^{\alpha_0}
= \frac{\nu-2}{\nu}
\left(
\left|
\mathbf{\Psi}^{-1}\widetilde{\boldsymbol{\Sigma}}_{\alpha_n} 
\right|
\right)^{1/p} .
\end{equation}
We approximate the unknown bias measure \eqref{eq:BiasDet} by its Monte Carlo estimate, say $\widehat{E}\{\widetilde{d}_{p,n,\nu}^{\alpha_0}\}$. We thus obtain
\begin{equation}\label{eq:biasdet}
    \delta_n =\left[\widehat{E}\{\widetilde{d}_{p,n,\nu}^{\alpha_0}\}\right]^{-1}
    =
    \left[
    \frac{1}{B}
    \sum_{b=1}^{B}
    \widetilde{d}_{p,n,\nu}^{\alpha_0,(b)}
    \right]^{-1} ,
\end{equation}
where $\widetilde{\boldsymbol{\Sigma}}_{\alpha_n}^{(b)}$ and $\widetilde{d}_{p,n,\nu}^{\alpha_0,(b)}$ are the MCD scatter estimate \eqref{eq:trimcov} and the related diagnostic \eqref{eq:detdia} computed from the $b$th Monte Carlo replication, while $B$ is the total number of such replications.
In this approach 
$\delta_n$ is a multiplicative correction applied to the estimated scatter matrix: if the determinant criterion is larger (smaller) than one, the corresponding
correction reduces (increases) the estimated scatter. By construction,
the target value is $\delta_n=1$ when the asymptotically calibrated
estimator is unbiased according to criterion \eqref{eq:detdia}.

The finite-sample calibration factor depends on the sample size and is expected to vary with the trimming proportion, the dimension and the tail heaviness of the underlying distribution.  Accordingly, we 
have
\[
\delta_n
=
\delta(n,p,\alpha_0,\nu) ,
\]
even if we prefer to keep the simpler notation $\delta_n$ 
whenever possible.
%
In our experiment $p$ and $\nu$ are fixed by design, while the dependence on $n$ and $\alpha_0$ is modelled empirically from Monte Carlo simulations.
Smooth regression functions are subsequently fitted to the simulated values in order to obtain calibration formulas suitable for routine implementation for any value of the input parameters in $\delta(n,p,\alpha_0,\nu)$.


\section{Overview of the Monte Carlo study}
\label{MC:sec}

Our Monte Carlo study has two main complementary objectives.
\begin{enumerate}
\item
First, we quantify the residual finite-sample discrepancy after applying the asymptotic factor \(\eta_{\alpha_0,p}(\nu)\) introduced in Section~\ref{Preliminaries:sec}. 
We then translate this residual finite-sample discrepancy into the determinant-based correction \(\delta_n\) introduced in Section \ref{calibration:sec}. The correction factor is saved from the full Monte Carlo experiment and then interpolated to form a set of calibration values computable for any quadriplet $(n,p,\alpha_0,\nu)$ within the investigated range.
\item
Second, we use the resulting scatter calibration to investigate whether it is sufficient for the main inferential quantities adopted in applications, i.e. the robust distances. This requires comparing the distance cutoff obtained with the asymptotic correction alone, say \(\kappa_{\rm RD}\), against the one obtained after additionally applying the determinant-based correction, say \(\kappa_{\rm RD}^{\delta}\); if a residual discrepancy remains, \(\kappa_{\rm RD}^{\delta}\) itself allows for a direct empirical cutoff correction.
\end{enumerate}
The first ojective belong to the scatter-calibration analysis investigated in Section
\ref{finite-sample:sec}, while distance-based calibration is developed in Section \ref{sec:distance_calibration}.
%
For the latter experiment, to avoid the circularity of evaluating a cutoff on the same observations used to estimate it, the analysis uses two separate sets of $B=1{,}000$ independent Monte Carlo replications: a first calibration set of $B$ replications, used to compute the empirical cutoff, and an independent evaluation set of $B$ replications, used solely to assess the resulting empirical size. 

The Monte Carlo study has been performed on a grid involving a wide range of sample sizes, dimensions, tail properties and trimming proportions:
\begin{eqnarray*}
n&\in&\{50,60,\ldots,500\},\quad \qquad \quad p\in\{2,3, \ldots, 8\},\\
\nu&\in&\{3,4, \ldots,10,15,20,30\},\quad \alpha_0\in\{0.05,0.10,\ldots,0.50\}.
\end{eqnarray*}
Nevertheless, we stress that our approach is easily replicable and extendable beyond our simulation grid through the code that we provide in our GitHub repository: see Section \ref{sec:simulation_implementation}.

All simulations were conducted in a MATLAB environment on a virtual machine running RedHat 8.10, with 32 cores and 64 GB of RAM. The MCD computations were performed using the \texttt{FSDA} implementation of the estimator (\url{https://www.fsdatoolbox.eu}), the only one presently addressing the case of uncontaminated data generated from a multivariate Student-\textit{t} distribution: see \texttt{modelT} option in \url{https://rosa.unipr.it/FSDA/mcd.html} and the related critical distance threshold value function in \url{https://rosa.unipr.it/FSDA/msdcutoff.html}.

 \section{Finite-sample correction of bias}
 \label{finite-sample:sec}

 \subsection{Methodology and algorithm}
 \label{sim_asymptotic_alg:sec}

The first experiment asks whether the asymptotic factor \(\eta_{\alpha_0,p}(\nu)\) given in Equation \eqref{eq:consistT} is sufficient to eliminate finite-sample scatter bias. The determinant-based criterion (\ref{eq:biasdet}) provides a natural measure of possible residual finite-sample discrepancy that remains after this asymptotic correction. 
Without loss of generality, thanks to affine equivarance of the MCD estimator, our target location vector is taken as
\[
\boldsymbol{\mu} = \boldsymbol{0},
\]
where $\boldsymbol{0}$ is a $p$-dimensional vector of zeroes, 
while the target covariance matrix is
\[
\boldsymbol{\Sigma}
=
\frac{\nu}{\nu-2}\mathbf{I}_p.
\]
where $\mathbf{I}_p$ is the $p$-dimensional identity matrix. 
Consequently, the diagnostic measure \eqref{eq:detdia} simplifies to
\[
\widetilde{d}_{p,n,\nu}^{\alpha_0}
= \frac{\nu-2}{\nu}
\left(
\left|
\widetilde{\boldsymbol{\Sigma}}_{\alpha_n}
\right|
\right)^{1/p} .
\]

To investigate the finite-sample scatter bias, 
we consider $B=1{,}000$ samples generated from a multivariate Student-\textit{t} distribution with $\nu>2$ degrees of freedom,
\[
\mathbf{X}_1^{(b)},\ldots,\mathbf{X}_n^{(b)} 
\sim
t_p(\mathbf{0},\mathbf{I}_p,\nu), \qquad b=1,\ldots,B.
\]
%
%
The raw MCD scatter estimator is computed for each simulated sample and corrected using the asymptotic Student-\textit{t} consistency factor
$\eta_{\alpha_0,p}(\nu)$. 
%
%
%
Hence, for the $b$th Monte Carlo replication, the diagnostic
%
\eqref{eq:detdia} is
\begin{equation}
\widetilde{d}_{p,n,\nu}^{\alpha_0,(b)}
=
\frac{\nu-2}{\nu}
\left|
\widetilde{\boldsymbol{\Sigma}}_{\alpha_n}^{(b)}
\right|^{1/p} .
\label{eq:biasdet_identity}
\end{equation}
The Monte Carlo estimate of bias to be plugged in criterion \eqref{eq:biasdet} is then 
\begin{equation}
d_{p,n,\nu}^{\alpha_0}
= \widehat{E}\{\widetilde{d}_{p,n,\nu}^{\alpha_0}\} =
\frac{1}{B}
\sum_{b=1}^{B}
\widetilde{d}_{p,n,\nu}^{\alpha_0,(b)}.
\label{eq:biasdet_mc}
\end{equation}
Values of $d_{p,n,\nu}^{\alpha_0}$ close to one thus indicate small residual finite-sample discrepancy according to the determinant criterion, whereas departures from one indicate residual bias after asymptotic calibration.
The complete procedure is summarized in Algorithm~\ref{alg:asymptotic_correction}.

\begin{algorithm}[t!]
\small 
\caption{\small Monte Carlo estimate of the finite-sample bias measure \eqref{eq:BiasDet} under a multivariate Student-\textit{t} model}
\label{alg:asymptotic_correction}
\vskip 2mm
\begin{algorithmic}[1]
\Require $n$, $p$, $\alpha_0$, $\nu>2$, number of replications $B$
\vskip 1mm
\Ensure Monte Carlo estimate $d_{p,n,\nu}^{\alpha_0}$
\State Compute the finite-sample trimming level $\alpha_n$ associated with $\alpha_0$, $n$ and $p$: see Equation~\eqref{eq:alphan}.
\State Compute the asymptotic Student-\textit{t} consistency factor
$\eta_{\alpha_0,p}(\nu)$ from Equation~\eqref{eq:consistT}.
\State Set
$\boldsymbol{\Sigma}=\frac{\nu}{\nu-2}\mathbf{I}_p$.
\For{$b=1,\ldots,B$}
    \State Generate
    $\mathbf{X}_1^{(b)},\ldots,\mathbf{X}_n^{(b)}
    \stackrel{\mathrm{i.i.d.}}{\sim}
    t_p(\mathbf{0},\mathbf{I}_p,\nu)$.
    \State Compute the raw MCD scatter estimator with trimming level $\alpha_n$.
    \State Apply the asymptotic consistency correction $\eta_{\alpha_0,p}(\nu)$ to obtain $\widetilde{\boldsymbol{\Sigma}}_{\alpha_n}^{(b)}$.
    \State Compute
    $\widetilde{d}_{p,n,\nu}^{\alpha_0,(b)}$
    according to Equation~\eqref{eq:biasdet_identity}.
\EndFor

\State Compute
$d_{p,n,\nu}^{\alpha_0}
=
B^{-1}\sum_{b=1}^{B}
\widetilde{d}_{p,n,\nu}^{\alpha_0,(b)}$ as in Equation \eqref{eq:biasdet_mc}.

\State \Return $d_{p,n,\nu}^{\alpha_0}$
\end{algorithmic}
\end{algorithm}

\subsection{Results}
\label{sim_asymptotic:sec}

Table~\ref{tab:determinant_calibration_summary} reports selected results from this simulation study. For each configuration, the table gives the Monte Carlo estimate of the determinant-based criterion after application of the asymptotic correction \(\eta_{\alpha_0,p}(\nu)\), its residual discrepancy from the target value, and the corresponding finite-sample calibration factor \(\delta_n\). The complete calibration results are retained in the calibration database described in Section~\ref{sec:simulation_implementation}.


The results show that the asymptotic consistency factor $\eta_{\alpha_0,p}(\nu)$ reduces the residual determinant bias, but does not completely eliminate it at finite sample sizes. 
For example, in the configuration $p=5$, $\alpha_0=0.5$, and $\nu=3$, the mean-root determinant criterion is $0.9686$ at $n=50$, corresponding to a residual bias of approximately $3.1\%$ on the determinant scale. 
The residual bias is most pronounced for small sample sizes and can vary appreciably with the dimension, degrees of freedom, and trimming parameter. For example, at $n=50$ and $p=5$, the estimated determinant-bias factor ranges from approximately $0.92$ to $0.97$ for the configurations reported in the table.

A peculiar result of the Student-\textit{t} model is a kind of ``blessing of heavy tails'', because the observed finite-sample bias increases with $\nu$ for any given dimension $p$ and trimming level $\alpha_0$. A similar finding was also observed by \citet[Table II]{cr+h:99} in terms of asymptotic efficiencies of the diagonal terms of the MCD estimator.
Although at present we do not have a theoretical explanation of this effect, we find its empirical appearance intriguing, and we plan to pursue its study in future research.

\begin{figure}
\centering\includegraphics[width=1\textwidth]{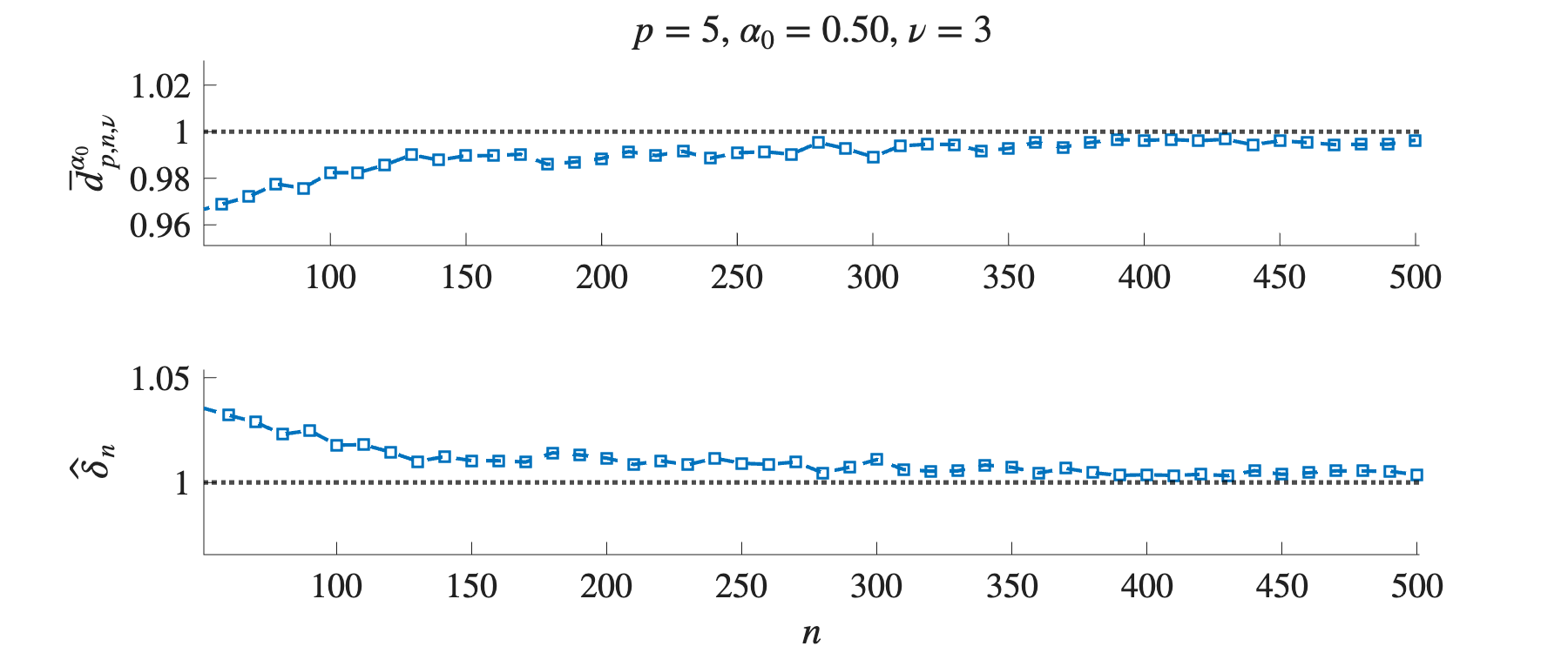}
\caption{\label{fig:determinant_calibration} finite-sample scatter discrepancy and resulting determinant-based calibration factor for $p=5$, $\alpha_0=0.50$ and $\nu=3$. The upper panel reports the Monte Carlo estimate $d_{p,n,\nu}^{\alpha_0}$ of the determinant-based criterion after application of the asymptotic Student-\textit{t} consistency factor $\eta_{\alpha_0,p}(\nu)$, with the horizontal dotted line denoting the target value one. The lower panel reports the corresponding finite-sample calibration factor $\delta_n=1/d_{p,n,\nu}^{\alpha_0}$. The correction is largest for small samples and approaches one as $n$ increases.}
\end{figure}

The discrepancy decreases as $n$ increases and becomes small for large samples, approaching the asymptotic target value $E\{\widetilde{d}_{p,n,\nu}^{\alpha_0}\}=1$ in \eqref{eq:BiasDet}.
Figure~\ref{fig:determinant_calibration} summarizes this behavior. The upper panel reports the Monte Carlo estimate \(d_{p,n,\nu}^{\alpha_0}\) of the determinant-based criterion, with the horizontal reference line at one. The systematic deviation below one for small and moderate sample sizes, followed by convergence toward one, provides direct evidence of a residual finite-sample effect after application of the asymptotic Student-\(t\) correction.
The magnitude of the discrepancy is therefore relevant primarily for small and moderate samples. This is consistent with the interpretation of \(\eta_{\alpha_0,p}(\nu)\) as an asymptotic consistency correction: it removes the population-level scaling induced by the Student-\(t\) model, but it does not account for all finite-sample effects associated with estimation of location and scatter and with the finite-sample MCD trimming level.
The residual discrepancies presented in this section provide the rationale for the finite-sample correction introduced in the subsequent Section \ref{sim_delta_pison:sec}.

\begin{table}[htbp]
\centering
\caption{Selected results from the Monte Carlo assessment of the 
finite-sample determinant calibration. The fifth column reports the Monte Carlo estimate $E\{\widetilde{d}_{p,n,\nu}^{\alpha_0}\}$ of the determinant-based criterion in Equation \eqref{eq:BiasDet} after application of the asymptotic correction $\eta_{\alpha_0,p}(\nu)$. The residual discrepancy is measured on the determinant scale as $100|E\{\widetilde{d}_{p,n,\nu}^{\alpha_0}\}-1|$, while $\delta_n$ 
is the corresponding finite-sample correction factor.
}
\label{tab:determinant_calibration_summary}
\begin{tabular}{rrrrrcr}
\hline \\
$p$ & $\nu$ & $n$ & $\alpha_0$ &
$E\{\widetilde{d}_{p,n,\nu}^{\alpha_0}\}$ & \% of residual  & $\delta_n$ \\
 &  &  &  &  & discrepancy &  \\
\hline
2 & 3 & 50  & 0.10 & 1.0044 & 0.44 & 0.9956 \\
2 & 3 & 50  & 0.50 & 0.9694 & 3.06 & 1.0315 \\
2 & 7 & 50  & 0.10 & 0.9893 & 1.07 & 1.0108 \\
2 & 7 & 50  & 0.50 & 0.9305 & 6.95 & 1.0747 \\
5 & 3 & 50  & 0.10 & 0.9317 & 6.83 & 1.0733 \\
5 & 3 & 50  & 0.50 & 0.9686 & 3.14 & 1.0324 \\
5 & 7 & 50  & 0.10 & 0.9364 & 6.36 & 1.0679 \\
5 & 7 & 50  & 0.50 & 0.9204 & 7.96 & 1.0865 \\
\hline
2 & 3 & 500 & 0.10 & 1.0014 & 0.14 & 0.9986 \\
2 & 3 & 500 & 0.50 & 0.9935 & 0.65 & 1.0066 \\
5 & 3 & 500 & 0.10 & 0.9898 & 1.02 & 1.0103 \\
5 & 3 & 500 & 0.50 & 0.9955 & 0.45 & 1.0045 \\
\hline
\end{tabular}
\end{table}

%



\begin{figure}[t!]
\centering
\begin{minipage}{\textwidth}
\centering
\includegraphics[width=0.65\textwidth]{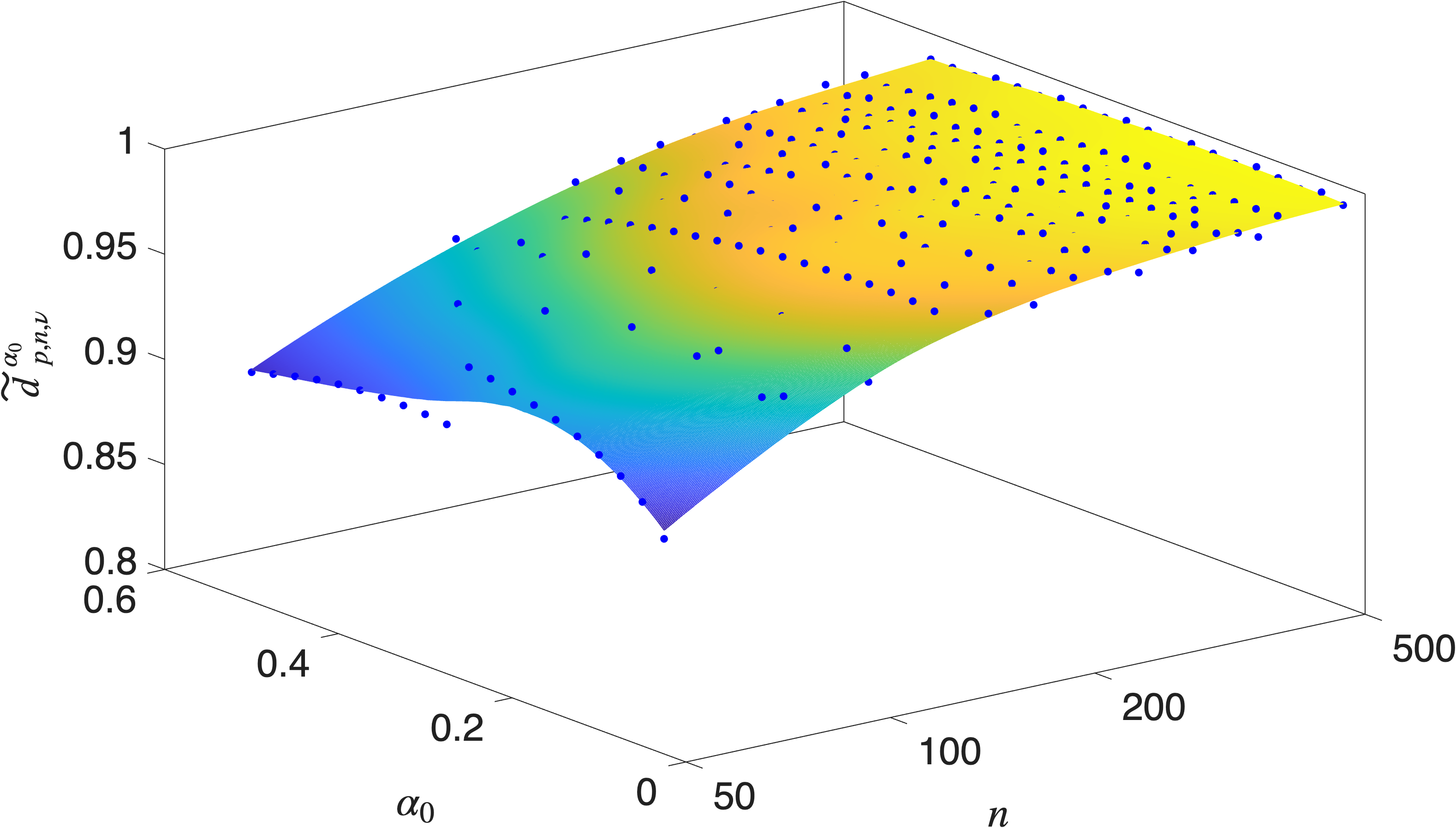}
\end{minipage}

\vspace{0.5cm}
\begin{minipage}{\textwidth}
\centering
\includegraphics[width=0.45\textwidth]{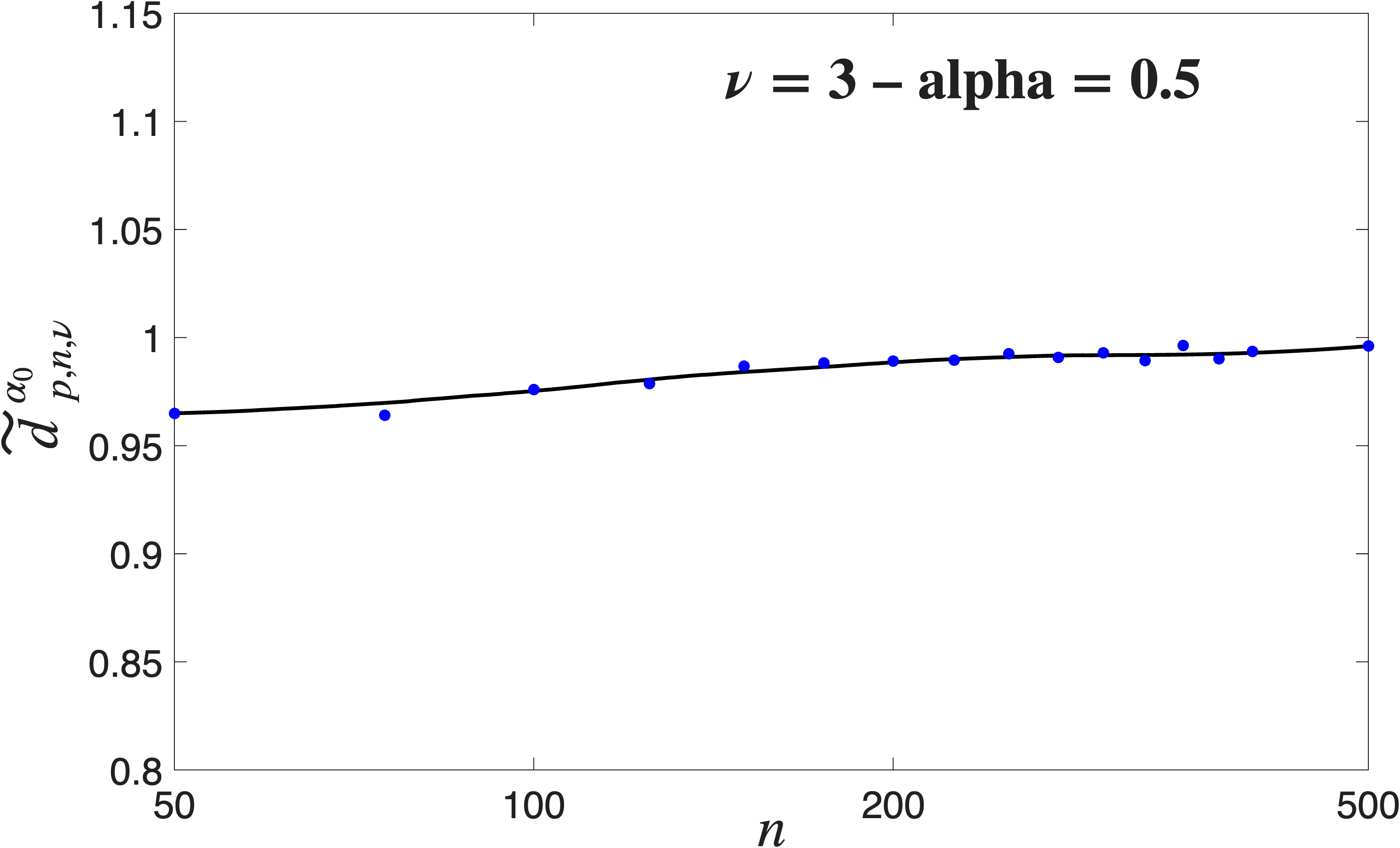}
\includegraphics[width=0.45\textwidth]{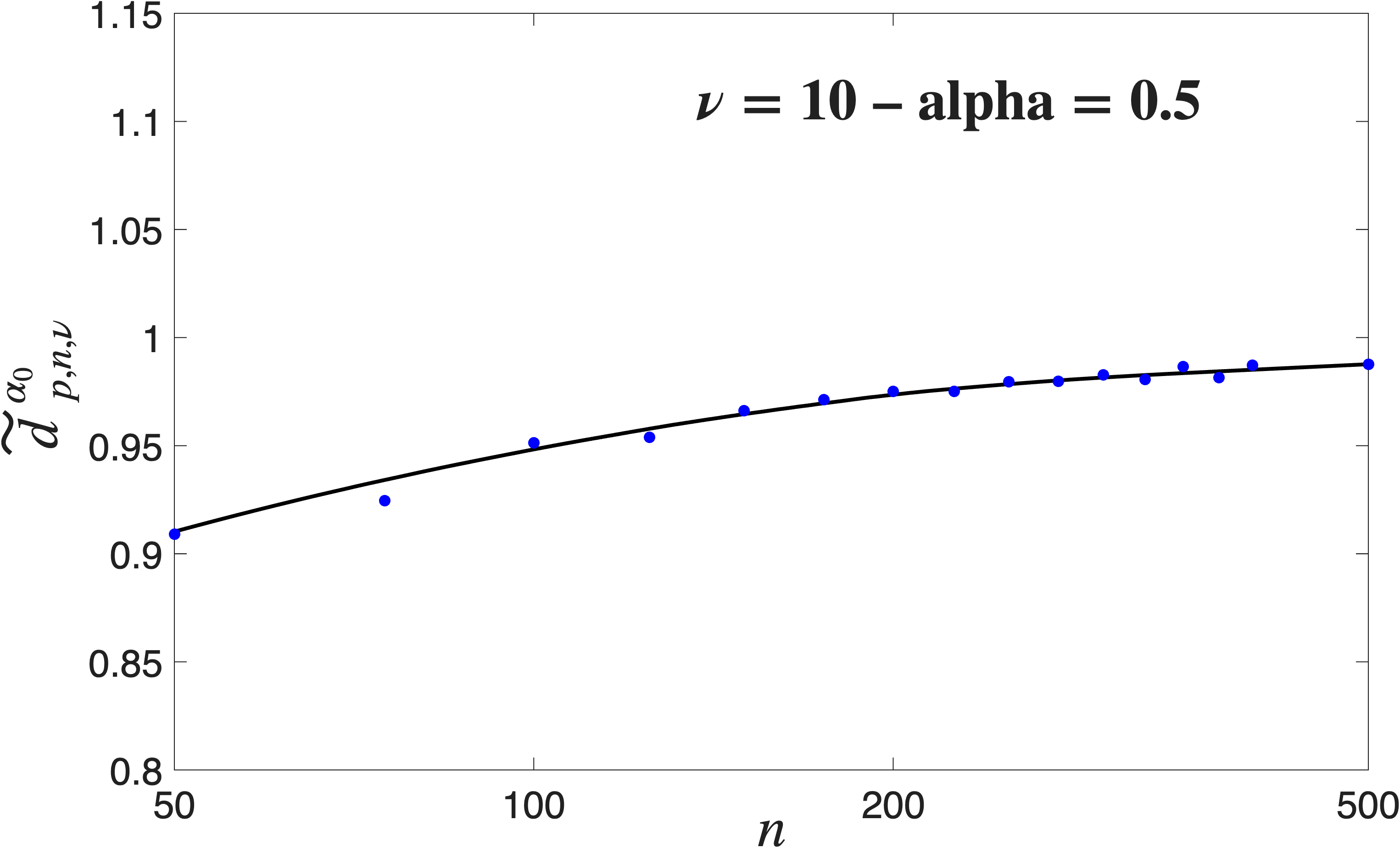}
\end{minipage}

\vspace{0.5cm}
\begin{minipage}{\textwidth}
\centering
\includegraphics[width=0.45\textwidth]{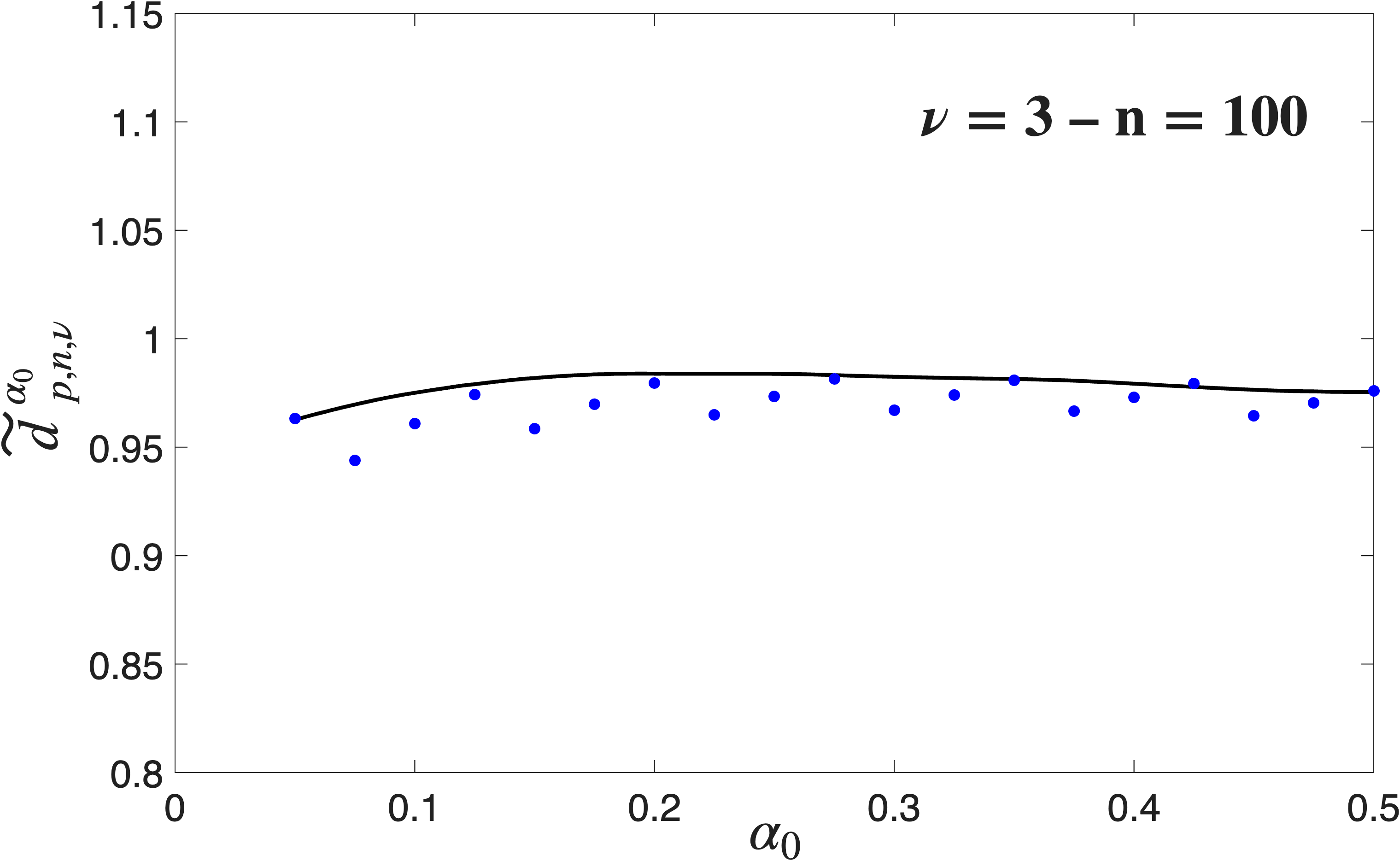}
\includegraphics[width=0.45\textwidth]{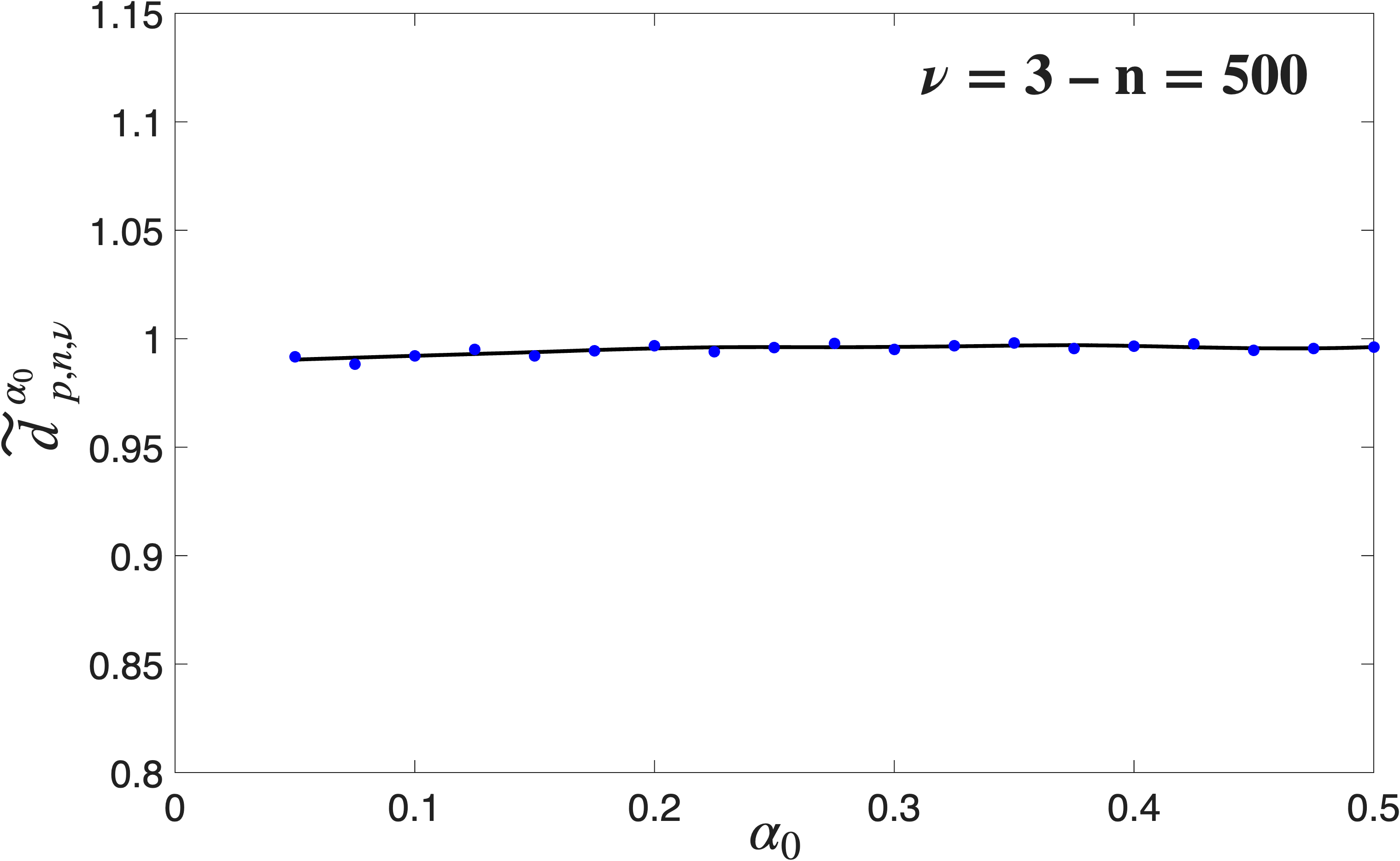}
\end{minipage}
\caption{Residual determinant bias diagnostic $\widetilde{d}_{p,n,\nu}^{\alpha_0}$ , after asymptotic Student-\textit{t} consistency correction, for $p=5$, and degrees of freedom $\nu=3$ as a function of sample size $n$, on a logarithmically spaced grid as in \cite{bar+al:26}, and $\alpha_0$ (top panel). 
Residual determinant bias diagnostic as a function of sample size $n$, when fixing $\alpha_0 = 0.5$ and for two different values of $\nu$ (central panels), and as a function of trimming level $\alpha_0$ when fixing $n$ to two different values ($100$ and $500$) for $\nu = 3$ (botton panels). }
\label{fig:bias_det}
\end{figure}

The origin of this finite-sample discrepancy was first observed during the broad-range simulations carried out in \citet{bar+al:26}, where high-breakdown MCD estimation was embedded into the theory of generalized radius processes to perform robust inference and outlier detection under contaminated Student-\textit{t} models. In that framework, the assessment of the empirical behavior of robust distances and conformance statistics required extensive Monte Carlo simulations spanning a wide spectrum of sample sizes, from small samples up to larger asymptotic regimes ($n = 2{,}000$).
%
The sample sizes in \cite{bar+al:26} were then selected on a relatively coarse and logarithmically spaced grid, with the aim of keeping the computation time feasible while providing a natural scale for smooth surface interpolation over orders of magnitude of $n$.
For $p=5$, Figure \ref{fig:bias_det} illustrates some complementary aspects that appear when merging the logarithmic grid of \cite{bar+al:26} with the specific target of the present paper:
\begin{itemize}
    \item The top panel shows the trajectory of $\widetilde{d}_{p,n,\nu}^{\alpha_0}$ as a function of the sample size and trimming proportion for the tail configuration $\nu = 3$.
    \item The central panels show the trajectory of $\widetilde{d}_{p,n,\nu}^{\alpha_0}$ as a function of the sample size under $\alpha_0=0.50$ for two tail configurations: $\nu = 3$ (heavy tails, left side) and $\nu = 10$ (lighter tails, right side). In both cases, the criterion approaches the asymptotic target value as $n$ increases, again with the finite-sample departure at $n=50$ being more pronounced for $\nu=10$ than for $\nu=3$. 
    \item The bottom panels display the behavior of $\widetilde{d}_{p,n,\nu}^{\alpha_0}$ as a function of the trimming proportion for $\nu=3$, comparing a small sample ($n=100$, left side) with a moderate sample ($n=500$, right side). The finite-sample discrepancy is clearly visible across all trimming levels when $n=100$, but largely attenuates by $n=500$. The $n=100$ panel also shows the effect of the finite-sample subset size of Equation (\ref{eq:hn}).
\end{itemize}
While these exploratory simulations clearly show the asymptotic validity of $\eta_{\alpha_0,p}(\nu)$ as $n \to \infty$, they also highlight that the bias remains non-negligible in the small-to-moderate sample regime ($n \le 500$). Because computing time is considerably reduced when attention is restricted to this region, the main body of this paper has considered a much denser and linearly spaced grid ($n \in \{50, 60, \dots, 500\}$) to construct the precise finite-sample calibration formulas.

\subsection{\mbox{Estimation of the finite-sample calibration factor}}
\label{sim_delta_pison:sec}

The determinant discrepancy estimated in Section~\ref{sim_asymptotic:sec} directly determines the additional finite-sample calibration factor. In particular, following \eqref{eq:deltacorrection}, for the mean-root-determinant criterion we have \(\widehat\delta_n=[d_{p,n,\nu}^{\alpha_0}]^{-1}\). Thus, the lower panel of Figure~\ref{fig:determinant_calibration} can be interpreted as the multiplicative correction required, in addition to \(\eta_{\alpha_0,p}(\nu)\), to recover the target determinant criterion.
For \(p=5\), \(\alpha_0=0.5\), and \(\nu=3\), the estimated correction is approximately \(1.03\) at \(n=50\). It decreases toward one with increasing sample size and is essentially negligible for \(n\) of the order of \(10^3\), as it should be. Hence, the additional determinant-based correction is mainly relevant for finite samples, while its asymptotic contribution vanishes as \(n\) increases.
The resulting values of \(\delta_n\) are retained from the Monte Carlo experiment and constitute the basis for the finite-sample scatter calibration used in the subsequent analysis.

\subsection{Interpolation of the calibration factors}
\label{sec:interpolation}

The Monte Carlo estimates $\widehat{\delta}_n$ of the finite-sample scatter calibration factor
\[
\delta_n=\delta(n,p,\alpha_0,\nu) ,
\]
evaluated at the configurations included in the simulation design, 
are the primary numerical results of the calibration experiment and are retained in the associated calibration database.
For practical implementation, a calibration value is also needed when the sample size or trimming proportion does not exactly coincide with a simulated configuration. For each fixed pair \((p,\nu)\), we therefore construct by interpolation a numerical representation of the simulated relation
\[
(n,\alpha_0)\longmapsto \widehat{\delta}(n,p,\alpha_0,\nu) = \big[ \widehat{d}^{\,\alpha_0}_{p,n,\nu} \big]^{-1}.
\]

The interpolation is performed following the simulation design for $(n,p,\alpha_0,\nu)$ of Section \ref{sim_asymptotic:sec}. 
%
The dimension \(p\) and the degrees of freedom \(\nu\) are treated as indexing parameters, whereas interpolation is performed over the two coordinates \(n\) and \(\alpha_0\). This avoids imposing an interpolation scheme over values of \(p\) or \(\nu\) that have not been directly investigated.
All evaluations are restricted to the simulated calibration domain; no extrapolation beyond that domain is used.
Nevertheless, as in Section \ref{MC:sec}, we stress that our approach is easily extendable beyond our simulation grid through the code that we provide in our GitHub repository.

The comparison of alternative representations is necessary because local variation in the simulated calibration factors can have two distinct sources for small samples. First, the factors are Monte Carlo estimates and hence may contain simulation variability. Second, the finite-sample MCD procedure depends on the integer-valued subset size \(h_n\), which changes deterministically with \((n,p,\alpha_0)\) according to \eqref{eq:hn}, and may induce genuine local features in the calibration surface. 

Exact piecewise-linear interpolation \citep{amidror:2002}, a standard bivariate scattered linear interpolation based on Delaunay triangulation, preserves both types of variation. LOESS \citep{cleveland:1979}, a two-dimen\-sional local linear surface smoother applied here with normalized coordinates and a span parameter of 0.15, yields a smoother calibration surface and may reduce Monte Carlo variability; however, it may also attenuate genuine variation associated with finite-sample MCD construction. The shape-preserving PCHIP representation \citep[Piecewise Cubic Hermite Interpolating Polynomials:][]{fritsch:1980,kahaner:1989} offers an intermediate approach, tracking local behavior in the simulated curves without the overshoot that can occur with unconstrained smooth interpolation. Consequently, we compare the three methods empirically rather than presupposing that local irregularities should be either preserved exactly or smoothed out.

\begin{figure}[tb!]
    \centering
    \includegraphics[width=1\linewidth]{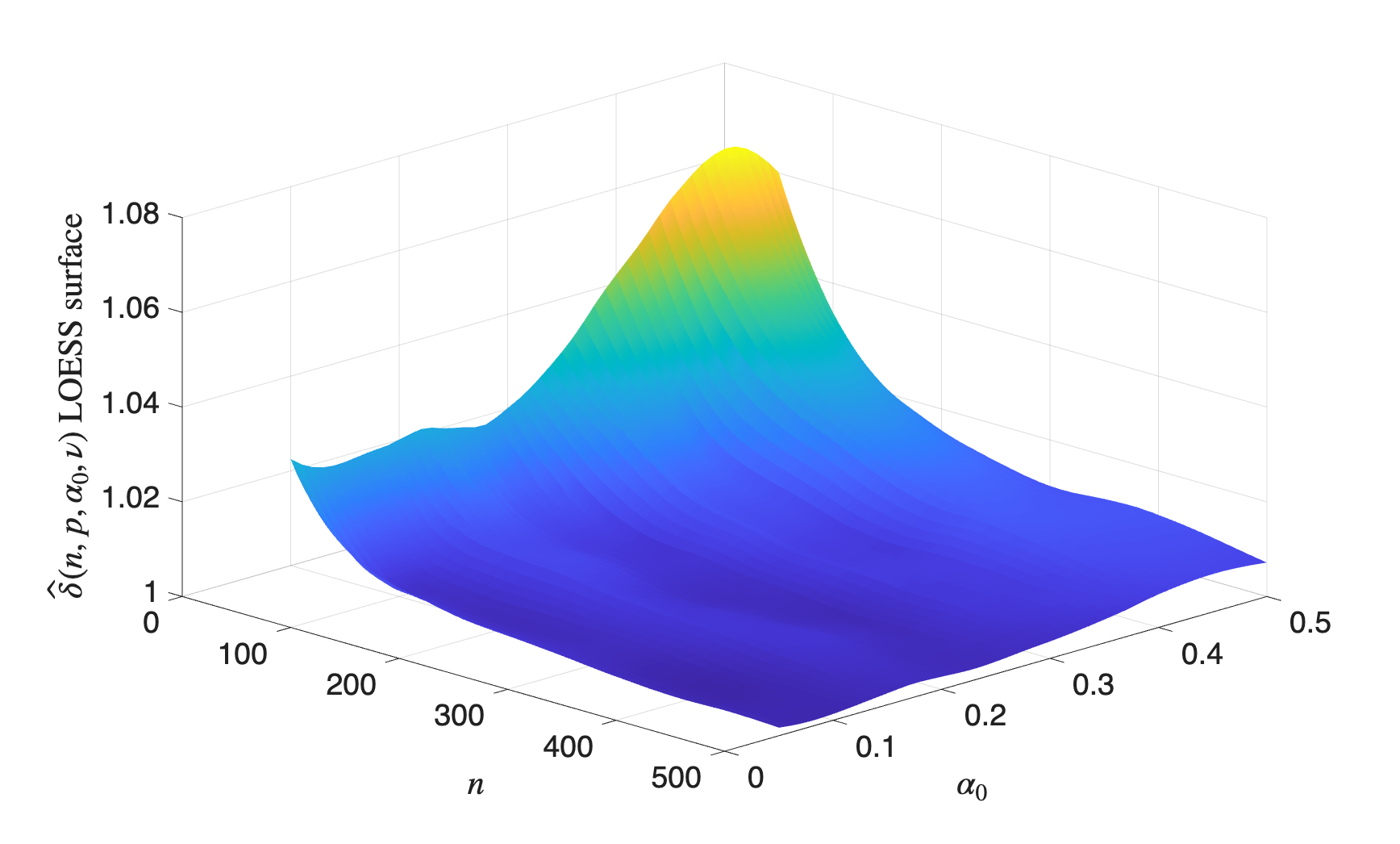}
    \caption{Interpolated values of 
$\widehat{\delta}(n,p,\alpha_0,\nu)$
in the case $p=2$ and $\nu=5$.}
    \label{fig:delta_loess_p2_nu5}
\end{figure}

Our Github repository (file \texttt{delta\_loess\_lookup.csv}, under folder \texttt{out\_article}) 
includes the full set of interpolated values of 
$\widehat{\delta}(n,p,\alpha_0,\nu)$ for the grid introduced in Section \ref{sim_asymptotic:sec} and using LOESS with a span of 0.15, which is the best performer according to the results in Table \ref{tab:delta_interpolation_validation}. Interpolation results are provided for virtually all sample sizes, dimensions, trimming levels and degrees of freedom of interest within our simulation range.
A pictorial representation of such results in the specific case $p=2$ and $\nu=5$ is given in Figure \ref{fig:delta_loess_p2_nu5}, as a function of $n$ and $\alpha_0$. 
The finite-sample MCD subset size $h_n$ and trimming level $\alpha_n$ are recorded for every configuration following their expressions \eqref{eq:hn} and \eqref{eq:alphan}. 
These quantities are deterministic functions of \((n,p,\alpha_0)\) and are not additional interpolation arguments.

We also use $h_n$ and $\alpha_n$ as diagnostics to interpret local features of the calibration curves. In particular, the analysis examines whether variations in the simulated factors are associated with changes in the integer-valued sequence \(h_n\).
The candidate representations are assessed through their residuals at the simulated configurations, their monotonicity and approach to the asymptotic value one, and their behavior under validation. For the latter, selected interior sample sizes are omitted when constructing the surface and their Monte Carlo calibration factors are compared with the corresponding predicted values. 

If $\delta_n^{l}$ denotes the Monte Carlo value of the calibration factor at the validation configuration $l$, and if $\hat{\delta}_n^{(l)}$ is its predicted value computed from a training sample that excludes configuration $l$, the prediction error is
\[
e_l = \delta_n^{(l)} - \hat{\delta}_n^{(l)}.
\]
We summarize in Table \ref{tab:delta_interpolation_validation} the validation performance over the $N_{\mathrm{val}}=6$ validation configurations described in detail in Appendix \ref{LOESS_interpolation_details:sec}. We compute the Mean Absolute Error 
and the Root Mean Squared Error 
\[
\mathrm{MAE} = \frac{1}{N_{\mathrm{val}}} \sum_{l=1}^{N_{\mathrm{val}}} |e_l| ,
\qquad \qquad
\mathrm{RMSE} = \left( \frac{1}{N_{\mathrm{val}}} \sum_{l=1}^{N_{\mathrm{val}}} e_l^2 \right)^{1/2} ,
\]
as our target measures. It is seen that LOESS clearly outperforms the other interpolation alternatives according to both criteria.

\begin{table}[t!]
\centering
\caption{Determinant-bias calibration factor $\widehat{\delta}$: summary of out-of-sample validation performance (RMSE and MAE) of candidate representations across dimensions $p \in \{2, 5\}$ and degrees of freedom $\nu \in \{3, 5, 7\}$ ($\alpha_0 \in [0.05, 0.50]$, $n \in [50, 500]$). Full details in Apendix \ref{LOESS_interpolation_details:sec}.}
\label{tab:delta_interpolation_validation}
\vspace{5mm}
\begin{tabular*}{\textwidth}{@{\extracolsep{\fill}} l cc @{}}
\toprule
\textbf{Candidate Method} & \textbf{Validation RMSE} & \textbf{Validation MAE} \\
\midrule 
\multicolumn{1}{@{}l}{\textbf{Overall Mean Across Surfaces:}} & & \\
\multicolumn{1}{@{}l}{\quad Linear Interpolation} & $0.00724$ & $0.00482$ \\
\multicolumn{1}{@{}l}{\quad Shape-preserving PCHIP} & $0.00781$ & $0.00515$ \\
\multicolumn{1}{@{}l}{\quad \textbf{LOESS ($\text{span}=0.15$)}} & $\mathbf{0.00448}$ & $\mathbf{0.00316}$ \\
\bottomrule
\end{tabular*}
\end{table}

\section{Distance-based finite-sample calibration}
\label{sec:distance_calibration}

Our second experiment investigates whether the residual finite-sample effects identified at the scatter level are also relevant for the robust distances used for outlier detection. 
The determinant-based factor $\delta_n$ introduced in Section~\ref{calibration:sec}, and its interpolated extension $\widehat{\delta}_n$, calibrates the scatter estimator with respect to a determinant criterion. This does not necessarily imply that the distribution of the resulting robust distances is correctly calibrated in finite samples. We therefore consider a separate correction acting directly on the cutoff used for robust-distance inference, following the general approach of \citet{cra:09}.
In contrast to \citet{cra:09}, who considered the MCD in the Gaussian case, the reference asymptotic distribution here is that of the (squared) robust distances under a multivariate Student-\textit{t} model.

As in the previous experiments, let
\[
\mathbf{X}_1,\ldots,\mathbf{X}_n
\stackrel{\mathrm{i.i.d.}}{\sim}
t_p(\mathbf{0},\mathbf{I}_p,\nu),
\qquad \nu>2,
\]
so that
\[
\boldsymbol{\Sigma}
=\frac{\nu}{\nu-2}\mathbf{I}_p.
\]
For a generic observation $\mathbf{X}$ from this population, define the squared Mahalanobis distance as
\[
D^2
=\mathbf{X}^{\top}\boldsymbol{\Sigma}^{-1}\mathbf{X}.
\]
Invariance of $D^2$ under affine transformations again ensures that choosing $\boldsymbol{\mu}=\boldsymbol{0}$ and $\boldsymbol{\Psi}=\mathbf{I}_p$ does not entail loss of generality.
\citet[Equation (12)]{bar+al:23} show that
\begin{equation}
\frac{1}{1+(\nu-2)/D^2}
\sim
\operatorname{Beta}\left(\frac{p}{2},\frac{\nu}{2}\right).
\label{eq:msd_beta}
\end{equation}
%
%
Hence, for a nominal left-tail probability $c\in(0,1)$, the cutoff for the \emph{squared} Mahalanobis distance $D^2$ under a multivariate Student-\textit{t} model is its $c$th quantile
\begin{equation}
q^2_{\nu,p}(c) = (\nu-2)\left[\frac{1}{1-I^{-1}_c(p/2,\nu/2)}-1\right],
\label{eq:student_cutoff}
\end{equation}
where, as in Equation \eqref{eq:consistT}, $I_u^{-1}(a,b)$ denotes the inverse of the regularized incomplete beta function. The corresponding cutoff on the natural (non-squared) distance scale is $q_{\nu,p}(c)=\sqrt{q^2_{\nu,p}(c)}$. 
We thus take $q^2_{\nu,p}(c)$ in Equation \eqref{eq:student_cutoff} as the asymptotic cutoff of the squared robust distances which are obtained by robustly estimating $\boldsymbol{\Sigma}$, provided that the selected estimator is consistent under the multivariate Student-\textit{t} model. It is well known that the MCD estimator satisfies this requirement \citep{cat+lop:12}.

In the simulations that follow, all finite-sample calibration is carried out on the squared scale, which is the scale returned by the \texttt{FSDA} function \texttt{msdcutoff.m} and by the corresponding main function \texttt{mcd.m}.
When appropriate, the natural-scale quantities are reported alongside for interpretability, but are not used in our calibration exercise. In the Gaussian limit $\nu\to\infty$, $q^2_{\nu,p}(c)$ converges to the usual $\chi^2_p$ cutoff.

For each Monte Carlo replication $b=1,\ldots,B$, let $\left(\widetilde{\boldsymbol{\mu}}_{\alpha_n}^{(b)},\widetilde{\boldsymbol{\Sigma}}_{\alpha_n}^{(b)}\right)$ denote the MCD location and scatter estimators obtained after applying the asymptotic consistency factor $\eta_{\alpha_0,p}(\nu)$. 
The squared robust distances are
\begin{equation}
{RD^2_i}^{(b)} = \left(\mathbf{X}_i^{(b)}-\widetilde{\boldsymbol{\mu}}_{\alpha_n}^{(b)}\right)^\top \left(\widetilde{\boldsymbol{\Sigma}}_{\alpha_n}^{(b)}\right)^{-1}\left(\mathbf{X}_i^{(b)}-\widetilde{\boldsymbol{\mu}}_{\alpha_n}^{(b)}\right),\quad i=1,\ldots,n,
\label{eq:RD_finite}
\end{equation}
with natural-scale counterpart $RD_i^{(b)}=\sqrt{{RD^2_i}^{(b)}}$.
For a fixed configuration $(n,p,\alpha_0,\nu)$, the empirical upper-tail cutoff in replication $b$ is defined as
\begin{equation}
q^{2,(b)}_{\mathrm{emp}}(c) = Q_c\left({RD^2_1}^{(b)},\ldots,{RD^2_n}^{(b)}\right),
\label{eq:qemp_rep}
\end{equation}
where $Q_c(\cdot)$ denotes the empirical quantile operator, which we computed following \cite{HyndmanYanan:1996} 
Type~7 estimator, with motivations detailed in Appendix \ref{quantiles:sec}. The Monte Carlo estimate of the finite-sample cutoff is then
\begin{equation}
q^2_{\mathrm{emp}}(c) = \frac{1}{B}\sum_{b=1}^B q^{2,(b)}_{\mathrm{emp}}(c),
\label{eq:qemp}
\end{equation}
with natural-scale counterpart $q_{\mathrm{emp}}(c)=\sqrt{q^2_{\mathrm{emp}}(c)}$. 
We set $B=1{.}000$ as in Section~\ref{finite-sample:sec}.

For extreme tail probabilities, or for very small samples, such that $(1-c)<1/n$, the distance-pooling method proposed by \cite{cra:09} can be adopted in place of \eqref{eq:qemp}. This method is also implemented in the code available through the GitHub site described in Section ~\ref{sec:simulation_implementation}.
Although we do not provide extensive evidence about the performance of the procedure of \cite{cra:09} in the present setting, our experience is that it generally provides results that are very similar to the computation of the Monte Carlo quantile estimate \eqref{eq:qemp}. 

The distance-based finite-sample calibration factor is defined as the ratio between the empirical and asymptotic cutoffs, both on the squared scale,
\begin{equation}
\kappa_{\mathrm{RD}}(n,p,\nu,\alpha_0,c) = \frac{q^2_{\mathrm{emp}}(c)}{q^2_{\nu,p}(c)}.
\label{eq:kappa_RD}
\end{equation}
Thus, $\kappa_{\mathrm{RD}}=1$ indicates exact agreement between the finite-sample and asymptotic cutoffs, whereas values different from one measure the residual finite-sample adjustment required for robust-distance inference.

%


The calibrated cutoff is obtained by scaling the theoretical threshold by this finite-sample adjustment factor:
\begin{equation}
q^2_{\mathrm{cal}}(c) = \kappa_{\mathrm{RD}}(n,p,\nu,\alpha_0,c)\, q^2_{\nu,p}(c),
\label{eq:qcal_RD}
\end{equation}
with natural-scale counterpart $q_{\mathrm{cal}}(c)=\sqrt{q^2_{\mathrm{cal}}(c)}$. While by construction $q^2_{\mathrm{cal}}(c) = q^2_{\mathrm{emp}}(c)$ at simulated grid points, $\kappa_{\mathrm{RD}}$ characterizes the magnitude of the finite-sample departure from asymptotic theory and enables smoothing or interpolation across configurations. 

The previous definitions use the asymptotically calibrated scatter estimator $\widetilde{\boldsymbol{\Sigma}}_{\alpha_n}^{(b)}$, and thus quantify the residual finite-sample discrepancy left by $\eta_{\alpha_0,p}(\nu)$ alone. In Section~\ref{calibration:sec} we introduce the doubly-calibrated estimator $\widehat{\boldsymbol{\Sigma}}_{\alpha_n}=\widehat\delta_n\widetilde{\boldsymbol{\Sigma}}_{\alpha_n}$. 
Replacing $\widetilde{\boldsymbol{\Sigma}}_{\alpha_n}^{(b)}$ with $\widehat{\boldsymbol{\Sigma}}_{\alpha_n}^{(b)}$ in Equation~\eqref{eq:RD_finite} yields squared corrected robust distances ${RD^2_i}^{(b),\delta}$, the replication-specific quantile $q^{2,(b),\delta}_{\mathrm{emp}}(c)$, and its Monte Carlo average $q^{2,\delta}_{\mathrm{emp}}(c)$, exactly as in 
\eqref{eq:qemp_rep}--\eqref{eq:qemp}. This defines a further calibration factor, again on the squared scale,
\begin{equation}
\kappa_{\mathrm{RD}}^{\delta}(n,p,\nu,\alpha_0,c) = \frac{q^{2,\delta}_{\mathrm{emp}}(c)}{q^2_{\nu,p}(c)},
\label{eq:kappa_RD_delta}
\end{equation}
with calibrated cutoff $q^{2,\delta}_{\mathrm{cal}}(c)=\kappa_{\mathrm{RD}}^{\delta}(n,p,\nu,\alpha_0,c)\,q^2_{\nu,p}(c)$ and natural-scale counterpart $q^{\delta}_{\mathrm{cal}}(c)=\sqrt{q^{2,\delta}_{\mathrm{cal}}(c)}$. 
Comparing $\kappa_{\mathrm{RD}}$ and $\kappa_{\mathrm{RD}}^{\delta}$ answers the question raised at the end of Section~\ref{sim_delta_pison:sec}: whether $\widehat\delta_n$ is, by itself, sufficient to calibrate the robust distances, or whether a further, distance-specific adjustment remains necessary.

The complete Monte Carlo procedure, evaluated for both $\widetilde{\boldsymbol{\Sigma}}_{\alpha_n}^{(b)}$ and $\widehat{\boldsymbol{\Sigma}}_{\alpha_n}^{(b)}$, is summarized in Algorithm~\ref{alg:distance_based_calibration}.
The (squared) cutoff of Equation \eqref{eq:student_cutoff} is computed 
with the \texttt{FSDA} function \texttt{msdcutoff.m}. 
In the Gaussian limit $\nu\rightarrow\infty$, 
the squared-distance cutoff converges to the usual $\chi^2_p$ cutoff, also covered as special case by function \texttt{msdcutoff.m} when $\nu$ is not provided or is conventionally set to 0.

\begin{algorithm}[t!]
\small
\caption{\small Monte Carlo estimate of the distance-based finite-sample calibration}
\label{alg:distance_based_calibration}
\vskip 2mm
\begin{algorithmic}[1]

\Require $n$, $p$, $\alpha_0$, $\nu>2$, number of replications $B$, confidence level $c$, finite-sample calibration factor $\widehat\delta_n$ from Section~\ref{sim_delta_pison:sec}

\Ensure $\kappa_{\mathrm{RD}}(n,p,\nu,\alpha_0,c)$ and $\kappa_{\mathrm{RD}}^{\delta}(n,p,\nu,\alpha_0,c)$

\State Compute the finite-sample trimming level $\alpha_n$ associated with $\alpha_0$, $n$ and $p$.

\For{$b=1,\ldots,B$}
    \State Generate $\mathbf{X}_1^{(b)},\ldots,\mathbf{X}_n^{(b)}
    \stackrel{\mathrm{i.i.d.}}{\sim} t_p(\mathbf{0},\mathbf{I}_p,\nu)$.
    \State Compute the MCD location and raw scatter using $\alpha_n$.
    \State Apply $\eta_{\alpha_0,p}(\nu)$ to the scatter estimator, obtaining $\widetilde{\boldsymbol{\Sigma}}_{\alpha_n}^{(b)}$.
    \State Compute the squared robust distances ${RD^2_i}^{(b)}$, $i=1,\ldots,n$, from $\widetilde{\boldsymbol{\Sigma}}_{\alpha_n}^{(b)}$, according to Eq. \eqref{eq:RD_finite}, and $q^{2,(b)}_{\mathrm{emp}}(c)$ according to Eq. \eqref{eq:qemp_rep}.
    \State Set $\widehat{\boldsymbol{\Sigma}}_{\alpha_n}^{(b)}=\widehat\delta_n\,\widetilde{\boldsymbol{\Sigma}}_{\alpha_n}^{(b)}$ and repeat the previous step with $\widehat{\boldsymbol{\Sigma}}_{\alpha_n}^{(b)}$ in place of $\widetilde{\boldsymbol{\Sigma}}_{\alpha_n}^{(b)}$, obtaining $q^{2,(b),\delta}_{\mathrm{emp}}(c)$.
\EndFor

\State Compute
$q^2_{\mathrm{emp}}(c) =
B^{-1}\sum_{b=1}^{B}q^{2,(b)}_{\mathrm{emp}}(c)$
and
$q^{2,\delta}_{\mathrm{emp}}(c) =
B^{-1}\sum_{b=1}^{B}q^{2,(b),\delta}_{\mathrm{emp}}(c)$.

\State Compute the asymptotic squared cutoff $q^2_{\nu,p}(c)$ from Eq. \eqref{eq:student_cutoff}.

\State Compute
\[
\kappa_{\mathrm{RD}}(n,p,\nu,\alpha_0,c)
=
\frac{q^2_{\mathrm{emp}}(c)}{q^2_{\nu,p}(c)},
\qquad
\kappa_{\mathrm{RD}}^{\delta}(n,p,\nu,\alpha_0,c)
=
\frac{q^{2,\delta}_{\mathrm{emp}}(c)}{q^2_{\nu,p}(c)}.
\]

\State \Return
$\kappa_{\mathrm{RD}}(n,p,\nu,\alpha_0,c)$ and $\kappa_{\mathrm{RD}}^{\delta}(n,p,\nu,\alpha_0,c)$.

\end{algorithmic}
\end{algorithm}

\begin{figure}
\centering\includegraphics[width=1\textwidth]{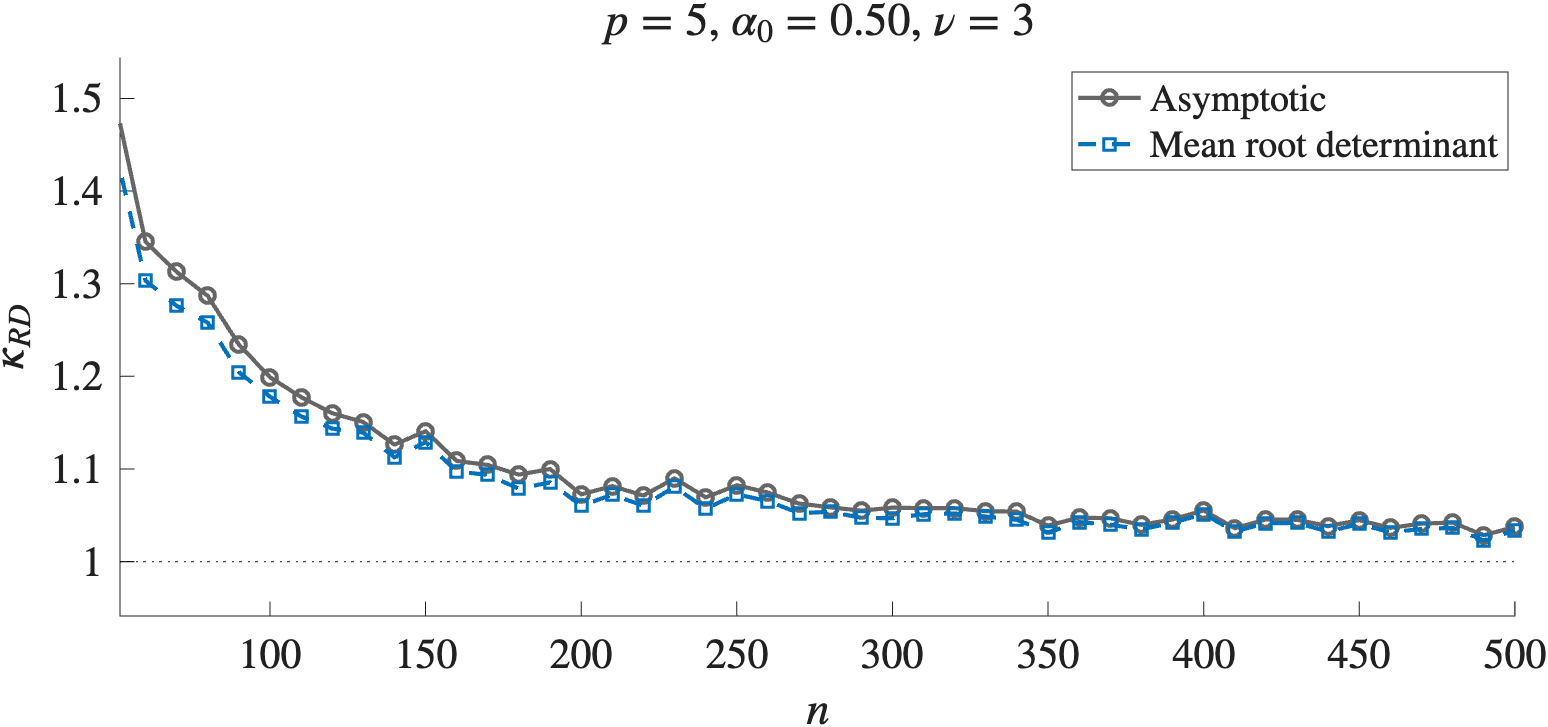}
\caption{\label{fig:kappa_RD}Distance-based finite-sample calibration for $p=5$, $\alpha_0=0.50$ and $\nu=3$. The solid grey curve reports $\kappa_{\rm RD}$ (Eq.~\eqref{eq:kappa_RD}), computed after applying only the asymptotic Student-\textit{t} consistency factor $\eta$ to the MCD scatter. The dashed blue curve reports $\kappa_{\rm RD}^{\delta}$ (Equation~\eqref{eq:kappa_RD_delta}), computed after additionally applying the determinant-based finite-sample correction $\widehat{\delta}_n$. The horizontal dotted line denotes the asymptotic target
$\kappa_{\rm RD}=1$. The determinant-based correction reduces the finite-sample discrepancy of the robust-distance cutoff, but does not eliminate it, especially for small samples.}
\end{figure}

Figure~\ref{fig:kappa_RD} shows the resulting values of $\kappa_{\mathrm{RD}}$ and $\kappa_{\mathrm{RD}}^{\delta}$, defined in Equations~\eqref{eq:kappa_RD} and \eqref{eq:kappa_RD_delta}, for $p=5$, $\alpha_0=0.5$ and $\nu=3$.
Using the asymptotically calibrated scatter alone ($\kappa_{\mathrm{RD}}$) requires a distance-cutoff adjustment of approximately \(1.51\) at \(n=50\). Additionally applying $\widehat\delta_n$ decreases the adjustment ($\kappa_{\mathrm{RD}}^{\delta}$) to approximately \(1.47\). 
Notice that these cutoff ratios are computed on the squared-distance scale used by \texttt{msdcutoff.m}. Consequently, since multiplication of the scatter matrix by $\widehat{\delta}_n$ divides the squared robust distances by $\widehat{\delta}_n$, we have
\[
\kappa_{\mathrm{RD}}^{\delta}
=
\frac{\kappa_{\mathrm{RD}}}{\widehat{\delta}_n} .  
\]
%
Thus, $\widehat\delta_n$ improves the calibration of the robust distances, but does not eliminate the finite-sample discrepancy.
The difference between $\kappa_{\mathrm{RD}}^{\delta}$ and $\kappa_{\mathrm{RD}}$ becomes progressively smaller as \(n\) increases. At \(n=1{,}000\), the corresponding values are approximately \(1.022\) and \(1.020\), respectively, indicating that both quantities approach the asymptotic 
value as expected.

Table~\ref{tab:RD_calibration} provides some concrete figures, showing that the determinant correction $\widehat\delta_n$ (which scales the scatter matrix by $\widehat{\delta}_{50} = \kappa_{\mathrm{RD}}/\kappa_{\mathrm{RD}}^{\delta} = 1.5094 / 1.4567 \approx 1.0362$, i.e., a $\approx 3.6\%$ correction) does not eliminate the finite-sample distortion in the distances. Indeed, $\kappa_{\mathrm{RD}}^{\delta}$ remains at $1.4567$ at $n=50$, representing a residual inflation of nearly $46\%$ over the asymptotic cutoff.
Thus, while correcting the scatter matrix is beneficial, the robust-distance distribution exhibits substantial additional finite-sample effects that are not captured by $\widehat\delta_n$ alone.
%
The fact that both $\kappa_{\mathrm{RD}}>1$ and $\kappa_{\mathrm{RD}}^{\delta}>1$ throughout the finite-sample range has a direct inferential interpretation: the asymptotic cutoff is systematically too low in finite samples, resulting in inflated empirical rejection rates (e.g., an empirical size of $8.50\%$ at $n=50$ and $6.43\%$ at $n=100$, compared to the nominal $\alpha=5\%$) when used without finite-sample distance calibration.

\begin{table}[t!]
\centering
\small
\caption{Distance-based finite-sample calibration for $p=5$, $\alpha_0=0.50$, $\nu=3$, and nominal significance level $0.05$ (asymptotic cutoff $q_{\nu,p}(0.95) = 3.8759$). For each sample size $n$, the table reports the squared-scale calibration factors ($\kappa_{\mathrm{RD}}$, $\kappa_{\mathrm{RD}}^{\delta}$) and the corresponding finite-sample cutoffs on the natural distance scale ($q_{\mathrm{emp}}$, $q_{\mathrm{emp}}^{\delta}$) under asymptotic consistency alone and under determinant double-calibration. The last column gives the empirical test size when the uncalibrated theoretical cutoff $q_{\nu,p}(0.95)$ is used.}
\label{tab:RD_calibration}
\begin{tabular}{r c cc c cc c c}
\toprule
& & \multicolumn{2}{c}{Consistency factor only ($\eta$)} & & \multicolumn{2}{c}{Doubly-calibrated ($\eta + \widehat{\delta}_n$)} & & Uncalibrated \\
\cmidrule(lr){3-4} \cmidrule(lr){6-7} \cmidrule(lr){9-9}
$n$ & & $\kappa_{\mathrm{RD}}$ & $q_{\mathrm{emp}}(c)$ & & $\kappa_{\mathrm{RD}}^{\delta}$ & $q_{\mathrm{emp}}^{\delta}(c)$ & & Size \\
\midrule
50   & & 1.5094 & 4.7619 & & 1.4567 & 4.6780 & & 0.08504 \\
100  & & 1.2039 & 4.2526 & & 1.1821 & 4.2141 & & 0.06427 \\
200  & & 1.1142 & 4.0912 & & 1.1082 & 4.0801 & & 0.05730 \\
500  & & 1.0365 & 3.9459 & & 1.0307 & 3.9349 & & 0.05252 \\
1000 & & 1.0234 & 3.9210 & & 1.0207 & 3.9158 & & 0.05157 \\
\bottomrule
\end{tabular}
\end{table}

\begin{figure}[t!]
\centering\includegraphics[width=1\textwidth]{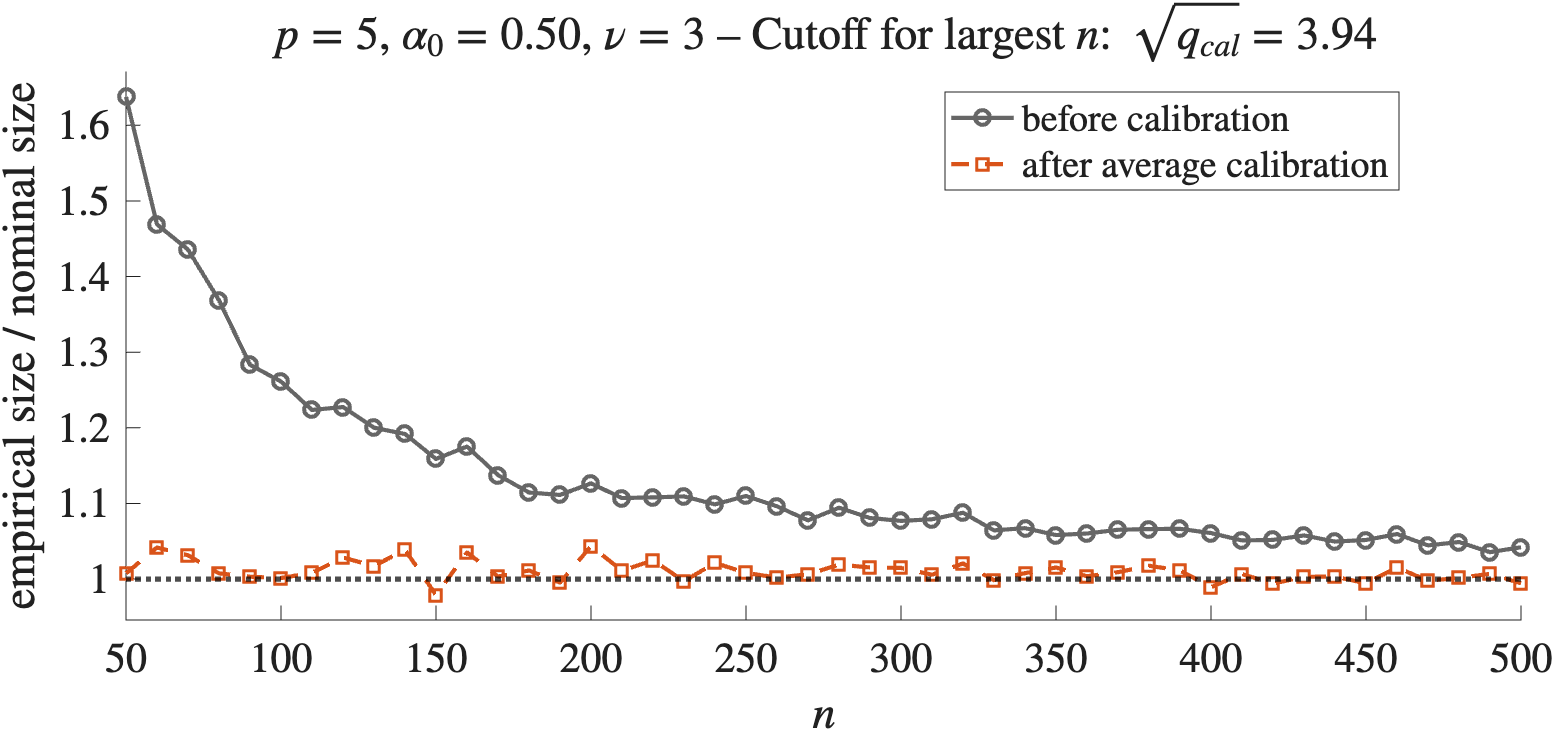}
\caption{\label{fig:test_size_calibration}Empirical outlier-test size, relative to the nominal significance level $0.05$, for $p=5$, $\alpha_0=0.50$ and $\nu=3$, using the determinant-corrected scatter $\widehat{\boldsymbol{\Sigma}}_{\alpha_n}=\widehat\delta_n\widetilde{\boldsymbol{\Sigma}}_{\alpha_n}$. The solid grey curve (``before calibration'') uses the asymptotic cutoff $q_{\nu,p}(c)$; the dashed blue curve (``after calibration'') uses the $\kappa_{\mathrm{RD}}^{\delta}$-calibrated cutoff $q_{\mathrm{cal}}^{\delta}(c)$ from Eq.~\eqref{eq:kappa_RD_delta}. The horizontal dotted line denotes the nominal size (ratio one). The distance-based calibration restores the correct test size at every sample size considered.}
\end{figure}

The practical effect of the distance-based correction is illustrated in Figure~\ref{fig:test_size_calibration}, which uses the determinant-corrected scatter $\widehat{\boldsymbol{\Sigma}}_{\alpha_n}$ throughout. The figure reports the empirical outlier-test size, relative to the nominal significance level, using the uncalibrated cutoff $q_{\nu,p}(c)$ (``before calibration'') and the $\kappa_{\mathrm{RD}}^{\delta}$-calibrated cutoff $q_{\mathrm{cal}}^{\delta}(c)$ (``after calibration''), for the same simulation setting.
Before calibration, the empirical test size is substantially above the nominal level for small samples, even though $\widehat\delta_n$ has already been applied to the scatter: at $n=50$ it is approximately $1.63$ times the nominal level, confirming that $\widehat\delta_n$ alone is not sufficient. The distance-based calibration $\kappa_{\mathrm{RD}}^{\delta}$ brings the empirical size back to the target value at every sample size considered. The discrepancy decreases progressively with $n$ even without the distance-based calibration, and the difference between the two procedures becomes negligible for large samples.

\begin{figure}[t!]
    \centering
    \includegraphics[width=1\linewidth]{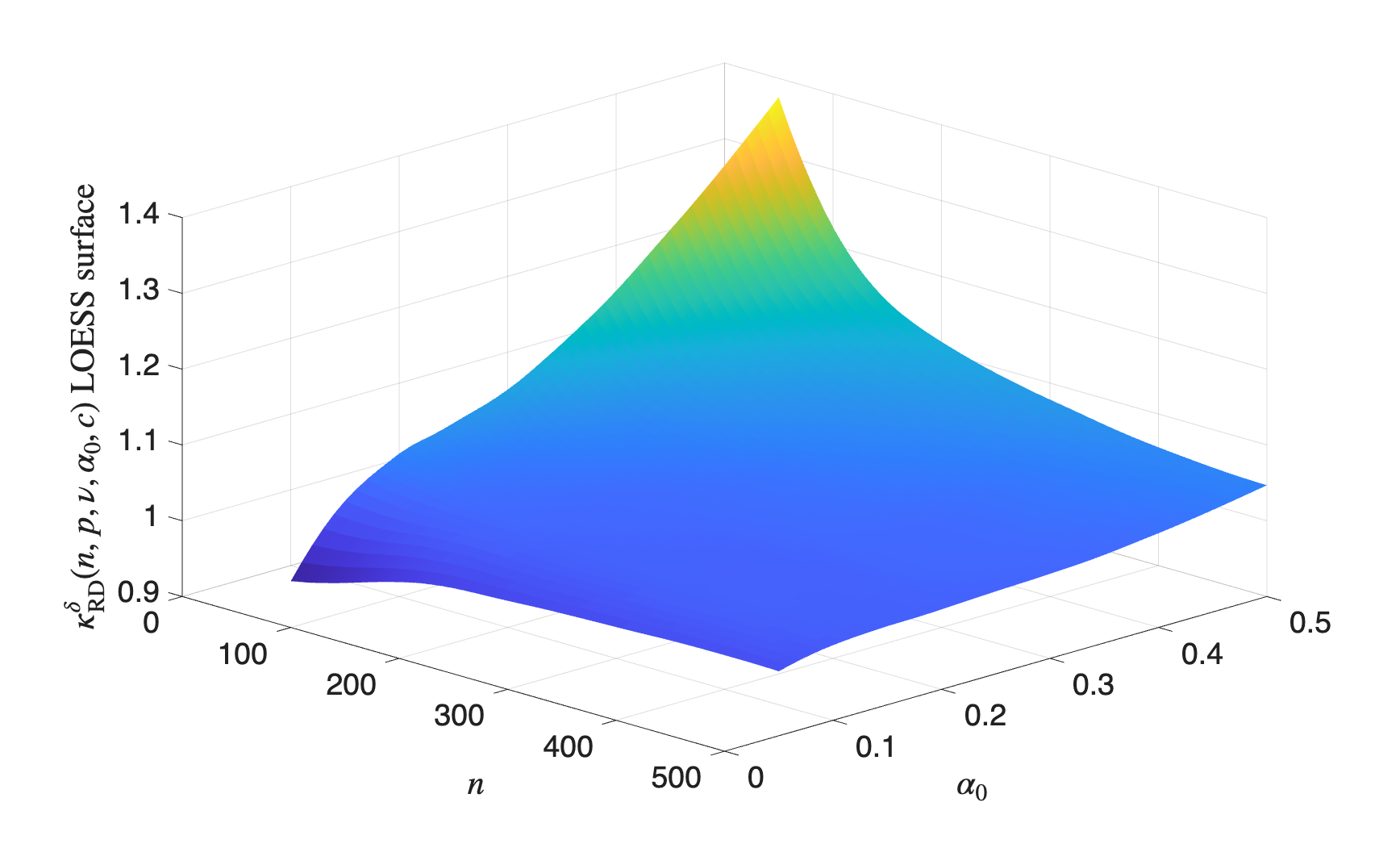}
    \caption{Interpolated values of $\widehat{\kappa}_{\mathrm{RD}}^{\delta}(n,p,\nu,\alpha_0,0.95)$ in the specific case $p=2$ and $\nu=5$.}
    \label{fig:RD_loess_p2_nu5_conf0.95}
\end{figure}

For practical application, continuous evaluations of $\kappa_{\mathrm{RD}}^{\delta}$ are obtained on the $(n, \alpha_0)$ plane for each fixed configuration of $(p, \nu, c)$ using the exact same three-candidate framework (Linear, LOESS, and PCHIP-in-$n$) and validation protocol formulated in Section~\ref{sec:interpolation}. This evaluation leads to the interpolated values $\widehat{\kappa}_{\mathrm{RD}}^{\delta}$, which  are also included in our Github repository (file \texttt{RD\_loess\_lookup.csv} under  folder \texttt{out\_article}). They are available for $c=0.95$ and for virtually all sample sizes, dimensions, trimming levels and degrees of freedom of interest within our simulation range.
Table~\ref{tab:RD_interpolation_validation} summarizes the out-of-sample validation performance of the tested interpolation approaches across the investigated combinations of $p$ and $\nu$ (see Appendix \ref{LOESS_interpolation_details:sec} for details). The superior performance of LOESS is apparent also in this scenario. Figure \ref{fig:RD_loess_p2_nu5_conf0.95} displays the interpolated 0.95-cutoffs in the specific case $p=2$ and $\nu=5$. Results for different cutoff probabilities can be produced in the same way and are not reported here. 
%
%
%

\begin{table}[t!]
\centering
\caption{Robust-distance cutoff calibration factor $\widehat{\kappa}_{\mathrm{RD}}^{\delta}(n,p,\nu,\alpha_0,c)$: summary of out-of-sample validation performance (RMSE and MAE) of candidate representations across dimensions $p \in \{2, 5\}$ and degrees of freedom $\nu \in \{3, 5, 7\}$ ($\alpha_0 \in [0.10, 0.50]$, $n \in [50, 500]$). Full details in Appendix \ref{LOESS_interpolation_details:sec}.}
\label{tab:RD_interpolation_validation}
\vspace{5mm}
\begin{tabular*}{\textwidth}{@{\extracolsep{\fill}} l cc @{}}
\toprule
\textbf{Candidate Method} & \textbf{Validation RMSE} & \textbf{Validation MAE} \\
\midrule
\multicolumn{1}{@{}l}{\textbf{Overall Mean Across Surfaces:}} & & \\
\multicolumn{1}{@{}l}{\quad Linear Interpolation} & $0.01856$ & $0.01017$ \\
\multicolumn{1}{@{}l}{\quad Shape-preserving PCHIP} & $0.01977$ & $0.01080$ \\
\multicolumn{1}{@{}l}{\quad \textbf{LOESS ($\text{span}=0.15$)}} & $\mathbf{0.01093}$ & $\mathbf{0.00727}$ \\
\bottomrule
\end{tabular*}
\end{table}

\section{Data analysis}
\label{data:sec}

\begin{figure}[t!]
\centering\includegraphics[width=0.9\textwidth]{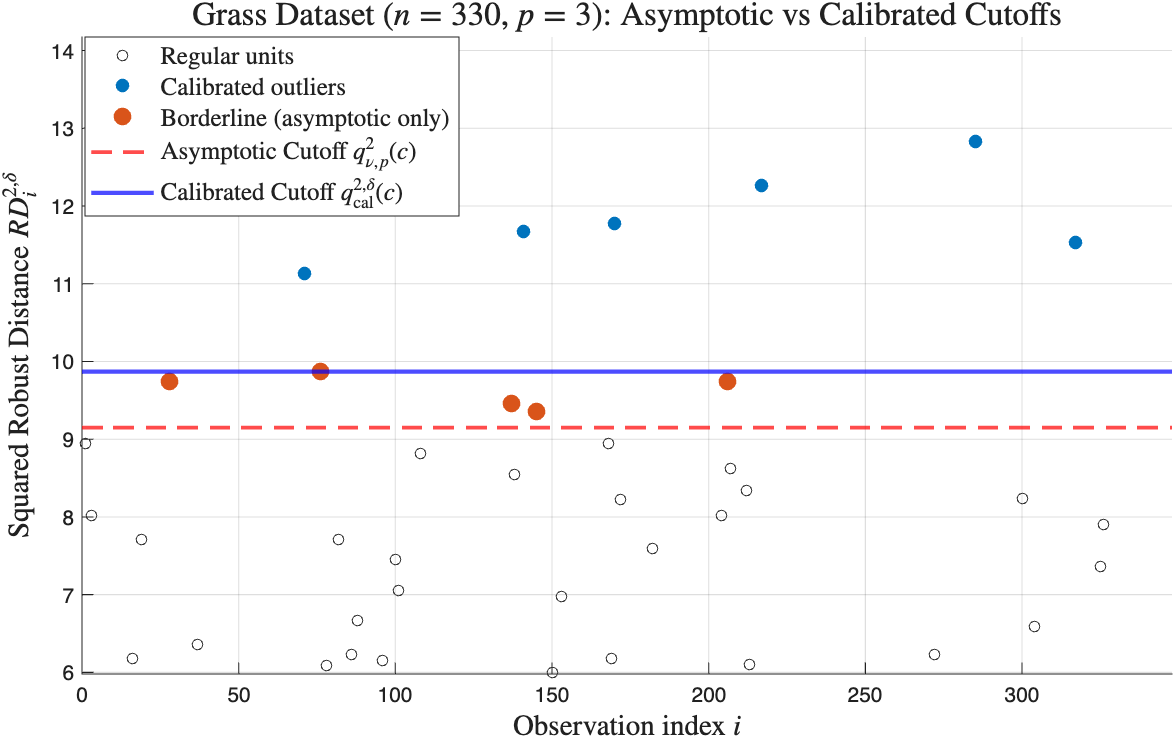}
\vspace{0.75cm}

\centering\includegraphics[width=0.9\textwidth]{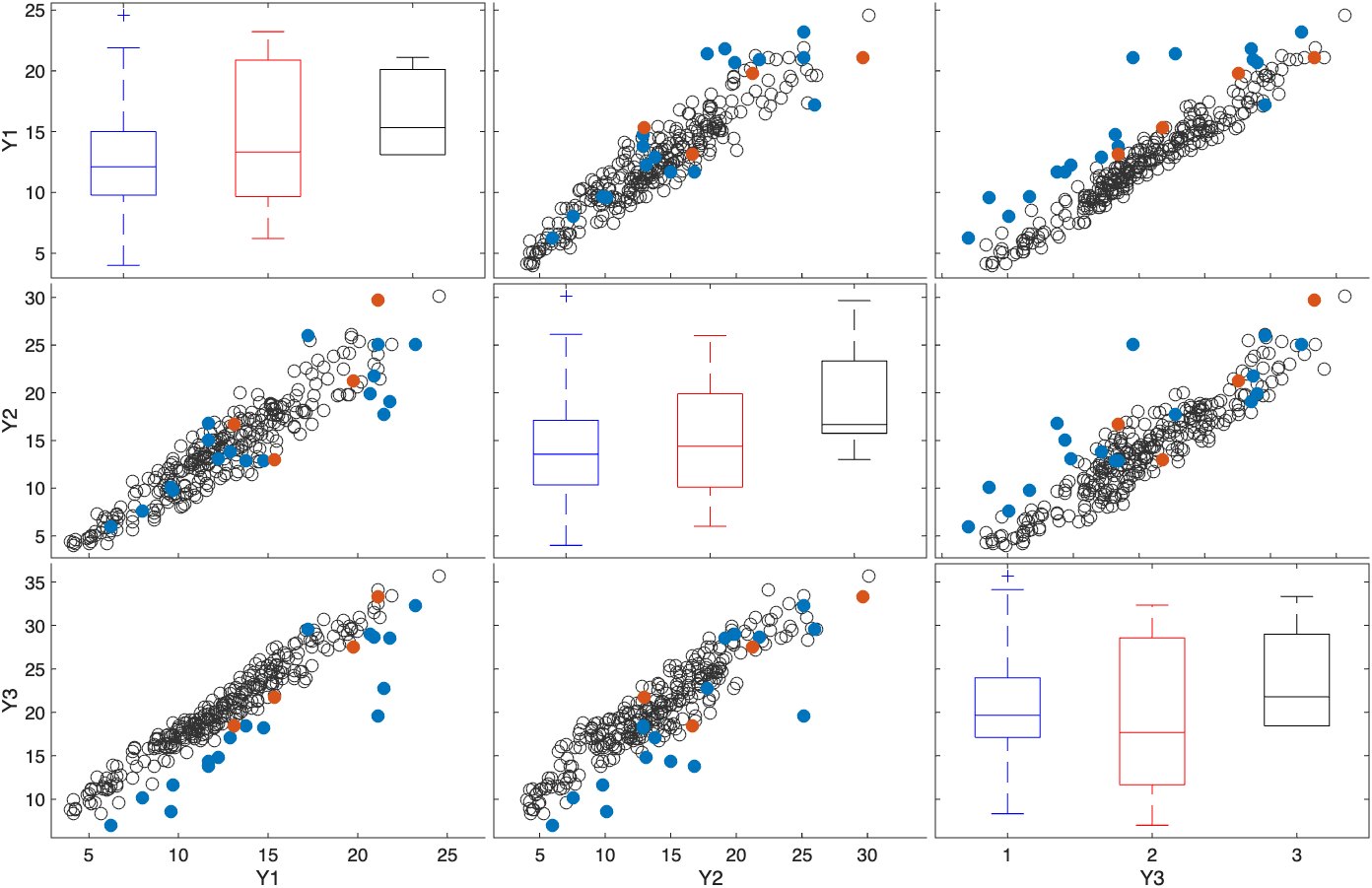}
\caption{\label{fig:Grass}Grass data, with $n=330$, $p=3$ and $\nu=8$. Upper panel: Zoom of the robust distance plot with $\alpha_0=0.5$. The dashed red line reports the asymptotic 0.95-quantile of the robust squared distances, while the solid blue line gives the corresponding calibrated quantile using \eqref{eq:kappa_RD_delta} and LOESS interpolation. Red circles denote the borderline pixels which are nominated outliers by the asymptotic threshold but not by our calibrated one.
Lower panel: Scatterplot matrix, where the same borderline pixels are represented with red circles.}
\end{figure}

We show the practical usefulness of our corrections by applying them to a problem already considered in \cite{bar+al:23} and \cite{bar+al:26}. Multivariate Student-\textit{t} models are gaining popularity in image analysis and in a benchmark data set for this task, available through the UCI Machine Learning Repository, 2,310 instances were randomly drawn
from a database of seven outdoor images. The images were
then hand-segmented to create a classification for every pixel.
Each instance is formed by a 3 × 3 pixel region and each
image is formed by $n=330$ instances. We analyze the image
labeled GRASS, which is supposed to be the most homogeneous
one in the whole data set. The absence of major contamination is important for our tasks, as we want to illustrate how accuracy in inference can be improved by the adoption of our corrections when a Student-\textit{t} model reasonably holds.

Our analysis focuses on the $p=3$ variables directly related to the classical RGB color decomposition. These variables provide the average of the Red, Green and Blue values, respectively, over each pixel
region. For these data, \cite{bar+al:26} robustly estimated 8 degrees of freedom for the hypothesized Student-\textit{t} data generating process, in good agreement with the value obtained using non-robust estimators for uncontaminated data. We thus take $\nu=8$ as a plausible value of the tail parameter and plug-in this value in all our corrections.

If we choose the asymptotic trimming proportion $\alpha_0=0.5$ that ensures maximal robustness, the asymptotic 0.95-cutoff for the squared robust distances is $q^2_{8,3}(0.95) = 9.15$. 
This cutoff is represented by the dashed red line in the upper panel of Figure \ref{fig:Grass}, which is a zoom of the plot of the $n=330$ squared robust MCD distances. This uncalibrated threshold yields 23 nominated outliers, a somewhat larger number than expected for $c=0.95$. Although such a deviation might be naively compatible with mild homogeneity in the image, we see that the application of our calibrated rule suggests that the asymptotic outlier test is too liberal even with a sample size as large as $n=330$. Indeed, the calibrated quantile using equation \eqref{eq:kappa_RD_delta} and LOESS interpolation, shown by the solid blue line in the upper panel of Figure \ref{fig:Grass}, does not label five bordeline units, depicted as red circles in the plot. The same units are also represented in the lower panel of Figure \ref{fig:Grass}, which shows the scatter plot matrix of the data and the full set of 18 nominated outliers after calibration. This number is now in much better agreement with the expected number of false positives for this uncantaminated example.

The code to reproduce these results is in Github (script \texttt{s03\_application.m}) together with a copy of the UCI dataset (under folder \texttt{datasets}).

\section{Implementation and reproducibility}
\label{sec:simulation_implementation}

%
%
Our findings can be replicated with codes available in GitHub at the address
\begin{center}
\url{https://github.com/UniprJRC/sscor-MCD}.    
\end{center}
For each configuration of the Monte Carlo study described in the preceding sections, the simulation code stores the determinant-based quantities required to construct the scatter calibration factor \(\widehat{\delta}_n\), together with the corresponding robust-distance calibration results. 

The numerical results are distributed as tabular calibration files. The file \texttt{BiasDet\_table\_mean\_root.csv} records the determinant-bias estimates indexed by \((\alpha_0,n,p,\nu)\), from which the scatter calibration factor is obtained. The file \texttt{mcd\_heavytails\_RD\_calibration.csv} records the robust-distance calibration results, additionally indexed by the confidence level and by the correction method. These files provide the numerical basis for the figures, the interpolation analysis, and the subsequent calibration formulas, independently of the simulation code.

The accompanying code reproduces the simulation design, constructs the calibration tables, computes the deterministic quantities \(q_n\), \(h_n\), and \(\alpha_n\), and performs the interpolation diagnostics. The code and calibration tables are made available in the project repository. 
Specifically, there are two main scripts needed to replicate the simulations, and one to guide to a concrete application to real datasets:
\begin{itemize}
    \item \texttt{s01\_mcdt\_finite\_sample\_calibration.m} runs the Monte Carlo simulation, estimates the determinant-bias corrections, calibrates the MCD robust-distance cutoff, evaluates empirical size, and writes the main calibration database;
    \item \texttt{s02\_mcdt\_interpolation\_analysis.m} consumes the simulation results and performs the interpolation/approximation analysis for both $\widehat{\delta}_n$ and $\kappa_{\mathrm{RD}}^{\delta}$.
    \item \texttt{s03\_application.m} demonstrates the application of the finite-sample MCD-t calibration,  after interpolation, to a real dataset, and compares the asymptotic cutoff with the calibrated cutoff. This is the practical step that makes the calibration usable in real applications, where the final sample size, dimension, tail parameter, and trimming level are not known in advance.
\end{itemize}

\noindent Our repository also provides the MATLAB code and the data necessary to replicate the main figures and tables of the manuscript. 
These codes can be used in two modalities:
\begin{itemize}
    \item in a standalone MATLAB licensed installation;
    \item in the license-free MATLAB Online (\url{https://matlab.mathworks.com/}).
\end{itemize}

The MCD estimator, which is at the core of the paper, and few other useful functions rely on the free MATLAB Add-On \texttt{FSDA} (see https://it.mathworks.com/matlabcentral/fileexchange/72999-fsda-flexible-statistics-data-analysis-toolbox). 
\texttt{FSDA} is a comprehensive suite designed for robust statistical analysis, supported by book-length documentation \citep{DataScienceWithMATLAB:2025,Robook:2024}.
This Add-On requires the Statistical and Machine Learning Toolbox. The LOESS computation has been performed using the Curve Fitting Toolbox. The Parallel Processing Toolbox is necessary if the reader needs to replicate the estimated corrections, which would take much more time otherwise.






\subsubsection*{Acknowledgements}
The work of Domenico Perrotta and Francesca Torti aligns
with the institutional program of the Joint Research Centre (JRC) of the
European Commission, under Horizon Europe (2021-2027). The \texttt{FSDA}
toolbox for \texttt{MATLAB} used in the simulations is a collaborative project between the University of Parma and the JRC. The authors gratefully acknowledge the tireless effort of Marco Riani for the development of such a project. The authors also thank Lucio Barabesi for helpful comments on a previous draft of this work.

\bibliography{Trimming_new3,Bib_new2}


\begin{appendices}

\section{Recall of the derivation of the determinant-based diagnostic  (\ref{eq:BiasDet_full})}
\label{recall_detdia:sec}

For a Student-\textit{t} distribution $X \sim t_p(\mu, \Psi, \nu)$ with $\nu > 2$, the true population covariance matrix is given by Equation \eqref{eq:covT}:
$$\Sigma = \frac{\nu}{\nu - 2} \Psi$$
Inverting $\Sigma$ yields:
$$\Sigma^{-1} = \left( \frac{\nu}{\nu - 2} \Psi \right)^{-1} = \left( \frac{\nu - 2}{\nu} \right) \Psi^{-1}$$
Substitute $\Sigma^{-1}$ directly into the determinant diagnostic (\ref{eq:detdia}):
\[
\tilde{d}_{p,n,\nu}^{\alpha_0} = \left| \Sigma^{-1} \tilde{\Sigma}_{\alpha_n} \right|^{1/p} = \left| \left( \frac{\nu - 2}{\nu} \right) \Psi^{-1} \tilde{\Sigma}_{\alpha_n} \right|^{1/p}
\]
Since $|k A| = k^p |A|$ for any scalar $k$ and $p\times p$ matrix $A$:
\[
\tilde{d}_{p,n,\nu}^{\alpha_0}  = \left[ \left( \frac{\nu - 2}{\nu} \right)^p \left| \Psi^{-1} \tilde{\Sigma}_{\alpha_n} \right| \right]^{1/p} \\
 = \left( \frac{\nu - 2}{\nu} \right) \left| \Psi^{-1} \tilde{\Sigma}_{\alpha_n} \right|^{1/p} .
\]

\section{Quantile estimation}
\label{quantiles:sec}

The Monte Carlo estimates of the empirical quantiles in Section \ref{sec:distance_calibration} rely on
MATLAB's \texttt{quantile} function, which supports several interpolation conventions for mapping the $n$ sorted observations to cumulative probabilities.
The default convention (\texttt{Method="midpoint"}, called \texttt{"exact"} prior to MATLAB R2025a) assigns the $k$-th sorted value to the cumulative probability $(2k-1)/(2n)$, corresponding to \cite{HyndmanYanan:1996} Type~5 estimator. An alternative convention, \texttt{Method="inclusive"}, assigns $(k-1)/(n-1)$, which forces the sample minimum and maximum to coincide exactly with the 0th and 100th percentiles, respectively. This is along Hyndman and Fan's Type~7, the long-standing default in R's \texttt{quantile()} for routine quantile estimation.

For the sample sizes 
considered here, the two conventions select interior order statistics that differ by only one or two ranks near $c=0.95$. Under the heavy-tailed models of interest, however, adjacent order statistics in the upper tail are widely spaced, so even this small shift can materially affect the resulting quantile estimate.
In fact, we found that the \texttt{"midpoint"} option introduces a systematic upward shift in $q_{\mathrm{emp,avg}}$ Eq.~\eqref{eq:kappa_RD_delta} for small $n$, causing the resulting calibrated cutoff to over-correct the empirical test size. In addition, with larger values of $c$, e.g., with $c=0.99$, the \texttt{"midpoint"} option would lead to select the observed largest value as the quantile estimate for the smallest sample sizes of our study.
Switching to \texttt{Method="inclusive"} removes such discrepancies, also bringing the average estimate \eqref{eq:qemp_rep} and the pooled calibration constructions of \cite{cra:09} into close agreement across the sample sizes considered.
For these reasons, in the present work all quantiles are computed in MATLAB with \texttt{Method="inclusive"}.

\begin{figure}[t!]
\begin{minipage}{\textwidth}
\centering\includegraphics[width=0.8\textwidth]{f3_test_size_calibration_B1000_inclusive.png}
\end{minipage} \\[5mm] 
\begin{minipage}{\textwidth}
\centering \includegraphics[width=0.9\textwidth]{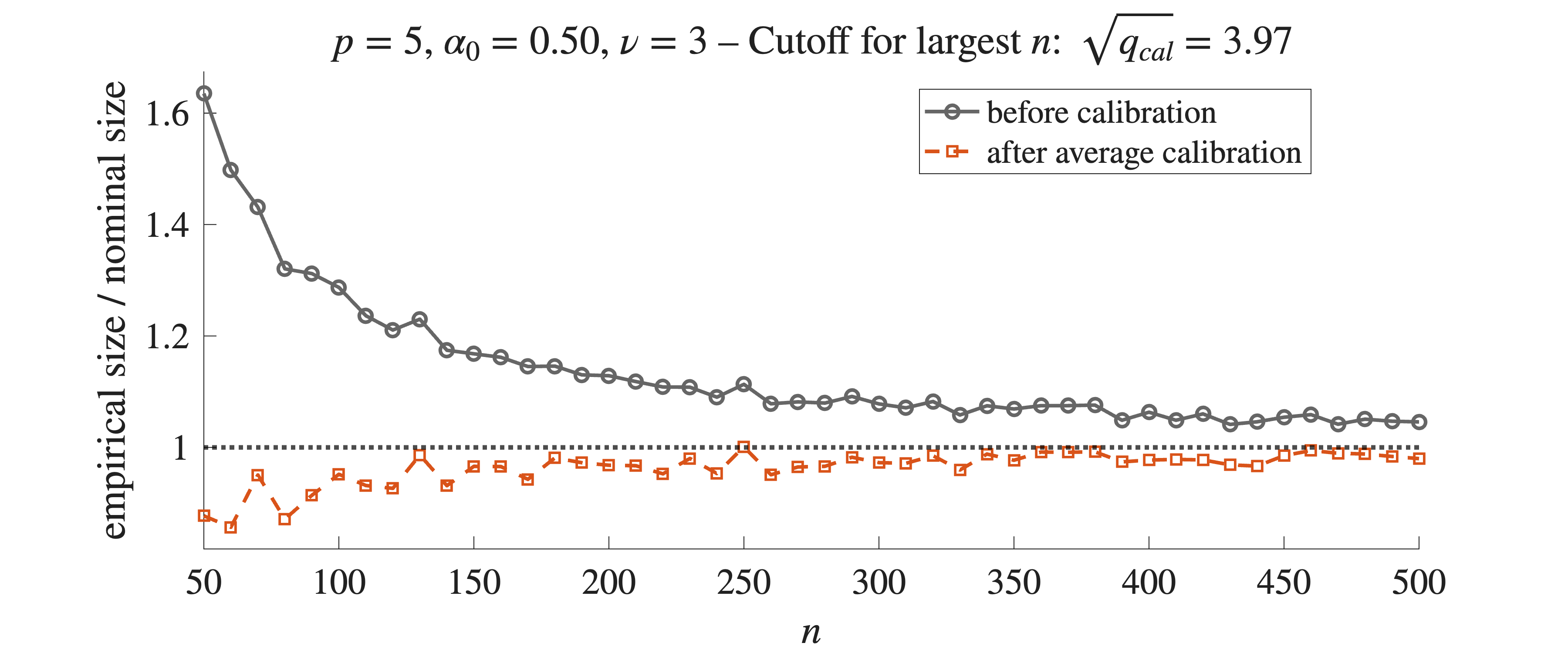}
\end{minipage} 
\caption{\label{ABC_inclusive_vs_midpoint:fig}
Calibration effect (with $B=1{,}000$ replicates) using MATLAB's \texttt{quantile} function with option \texttt{Method="inclusive"} (top panel, as in Figure \ref{fig:test_size_calibration}, re-proposed here for clarity) and option \texttt{Method="midpoint"} (bottom panel).}
\end{figure}

To assess the practical relevance of this choice, we compared the calibration results obtained with the two conventions. Figure~\ref{ABC_inclusive_vs_midpoint:fig} reports at the top the results obtained with \texttt{Method="inclusive"}, whereas the bottom panel shows the corresponding results obtained with MATLAB's default \texttt{Method="midpoint"}. The difference is visible, particularly for the smaller sample sizes. 
With the \texttt{"midpoint"} convention, the empirical size after average calibration is systematically below the nominal value for small $n$, indicating that the resulting calibration cutoff is too conservative. In contrast, the use of \texttt{"inclusive"} substantially reduces this discrepancy and keeps the calibrated empirical size much closer to the nominal level.

%
The difference is also reflected directly in the estimated calibration
cutoff. For the setting
illustrated in the two figures, the cutoff obtained with
\texttt{Method="inclusive"} for $n=500$ (reported in the title of each panel) is
\[
\sqrt{q_{\mathrm{cal}}}=3.94,
\]
whereas the corresponding value obtained with
\texttt{Method="midpoint"} is
\[
\sqrt{q_{\mathrm{cal}}}=3.97.
\]
The corresponding values for $n=50$ (not reported in the panel but available in the results table) are $4.59$ and $4.96$,
respectively. Thus, the choice of quantile convention has a
substantial effect on the calibration cutoff for small samples:
the \texttt{"midpoint"} convention produces a noticeably larger
cutoff, and hence a more conservative test, with the difference
becoming much smaller as $n$ increases. This is consistent with the
empirical-size results in the panels of Figure~\ref{ABC_inclusive_vs_midpoint:fig}, where the use of \texttt{"midpoint"} leads
to a more pronounced downward deviation from the nominal test size
for small $n$, whereas the two conventions give increasingly similar
results as the sample size grows.

For completeness, the choice between \texttt{"inclusive"} and \texttt{"midpoint"} should be distinguished from MATLAB's \texttt{Method="approximate"}, which uses a T-digest representation of the empirical distribution. The comparisons above use the sorting-based quantile calculation and therefore concern only the finite-sample interpolation convention, not the additional approximation introduced by a T-digest.

This comparison is also useful for interpreting the different calibration constructions.
\begin{figure}
\centering\includegraphics[width=0.8\textwidth]{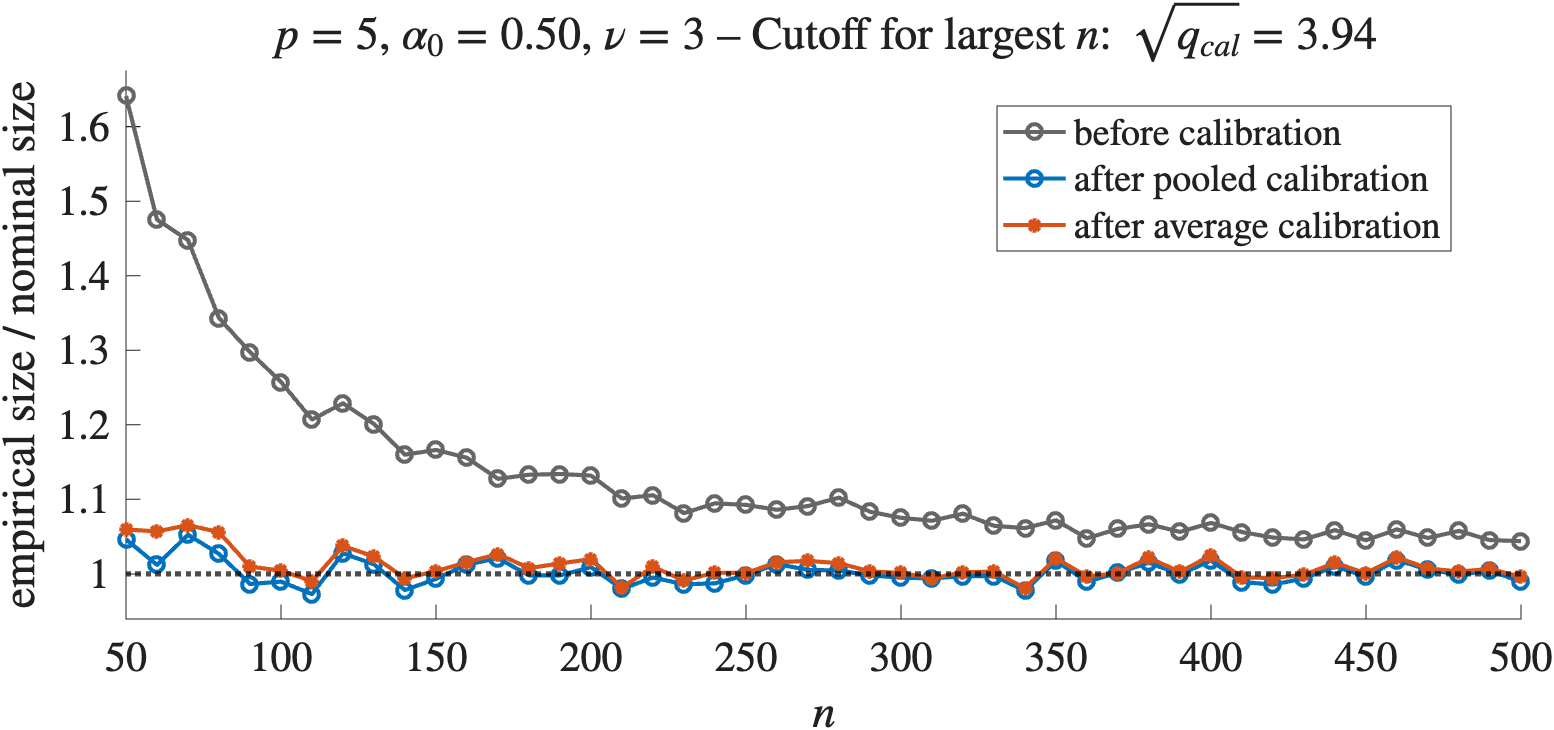}
\caption{\label{ABC_inclusive:fig}Comparison of the pooled calibration method of \cite{cra:09} (A) and and the average calibration approach adopted in \eqref{eq:qemp} (B), with $B=1{,}000$, using MATLAB's \texttt{quantile} function with option \texttt{Method=`inclusive'}.}
\end{figure}
Figure \ref{ABC_inclusive:fig} shows, with \texttt{Method="inclusive"}, that the pooled calibration method of \cite{cra:09} (A) and the average calibration approach adopted in \eqref{eq:qemp} (B) are closely aligned over the range of sample sizes considered, supporting their interpretation as two implementations of the same common-calibration principle.

\section{Out-of-sample validation and selection of the LOESS interpolant}
\label{LOESS_interpolation_details:sec}

This appendix provides the configuration-by-configuration validation results comparing the three candidate representations introduced in Section~\ref{sec:interpolation} and also exploited in Section \ref{sec:distance_calibration}: exact piecewise-linear interpolation, two-dimensional LOESS (with span $\alpha_{\mathrm{span}} = 0.15$), and the hybrid PCHIP (in-$n$/linear-in-$\alpha_0$) method. Predictive accuracy is assessed through out-of-sample validation by omitting interior sample sizes $n_{\mathrm{val}} \in \{70, 120, 170, \dots, 470\}$ during surface estimation and computing the Root Mean Squared Error (RMSE) and Mean Absolute Error (MAE) at the withheld nodes. 

Table~\ref{tab:delta_interpolation_validation_full} details the performance across all simulated $(p, \nu)$ surfaces for the determinant-based scatter calibration factor $\widehat{\delta}(n,p,\alpha_0,\nu)$, while Table~\ref{tab:RD_interpolation_validation_full} reports the corresponding behavior across $(p, \nu, c)$ configurations for the robust-distance cutoff calibration factor $\widehat{\kappa}_{\mathrm{RD}}^{\delta}(n,p,\nu,\alpha_0,c)$. Summaries of these results are given in Table~\ref{tab:delta_interpolation_validation} and in Table~\ref{tab:RD_interpolation_validation}, respectively

Across all dimensions and degrees of freedom, the LOESS representation systematically achieves the lowest prediction errors, reducing average RMSE and MAE by over 38\% and 34\% for $\widehat{\delta}$ (and by approximately 41\% and 28\% for $\widehat{\kappa}_{\mathrm{RD}}^{\delta}$) relative to piecewise-linear interpolation. This confirms that local polynomial smoothing effectively filters out finite-sample Monte Carlo variability without introducing noticeable boundary distortion. Consequently, the LOESS surface has been adopted throughout the paper as the definitive interpolation method.

\begin{table}[t!]
\centering
\small
\caption{\textbf{Determinant-bias calibration factor} $\widehat{\delta}_n$: out-of-sample validation performance (RMSE and MAE) of candidate representations across dimensions $p \in \{2, 5\}$ and degrees of freedom $\nu \in \{3, 5, 7\}$ ($\alpha_0 \in [0.05, 0.50]$, $n \in [50, 500]$).}
\label{tab:delta_interpolation_validation_full}
\footnotesize
\begin{tabular*}{\textwidth}{@{\extracolsep{\fill}} ccl cc @{}}
\toprule
$p$ & $\nu$ & \textbf{Candidate Method} & \textbf{Validation RMSE} & \textbf{Validation MAE} \\
\midrule
$2$ & $3$ & Linear Interpolation & $0.00907$ & $0.00608$ \\
    &     & Shape-preserving PCHIP & $0.00964$ & $0.00646$ \\
    &     & LOESS & $\mathbf{0.00592}$ & $\mathbf{0.00416}$ \\
\addlinespace[0.3em]
$2$ & $5$ & Linear Interpolation & $0.00801$ & $0.00544$ \\
    &     & Shape-preserving PCHIP & $0.00848$ & $0.00572$ \\
    &     & LOESS & $\mathbf{0.00513}$ & $\mathbf{0.00360}$ \\
\addlinespace[0.3em]
$2$ & $7$ & Linear Interpolation & $0.00750$ & $0.00521$ \\
    &     & Shape-preserving PCHIP & $0.00812$ & $0.00556$ \\
    &     & LOESS & $\mathbf{0.00478}$ & $\mathbf{0.00355}$ \\
\midrule
$5$ & $3$ & Linear Interpolation & $0.00744$ & $0.00464$ \\
    &     & Shape-preserving PCHIP & $0.00791$ & $0.00495$ \\
    &     & LOESS & $\mathbf{0.00411}$ & $\mathbf{0.00281}$ \\
\addlinespace[0.3em]
$5$ & $5$ & Linear Interpolation & $0.00593$ & $0.00392$ \\
    &     & Shape-preserving PCHIP & $0.00653$ & $0.00421$ \\
    &     & LOESS & $\mathbf{0.00353}$ & $\mathbf{0.00247}$ \\
\addlinespace[0.3em]
$5$ & $7$ & Linear Interpolation & $0.00551$ & $0.00365$ \\
    &     & Shape-preserving PCHIP & $0.00619$ & $0.00399$ \\
    &     & LOESS & $\mathbf{0.00343}$ & $\mathbf{0.00237}$ \\
\bottomrule
\end{tabular*}
\end{table}

\begin{table}[t!]
\centering
\small
\caption{\textbf{Robust-distance cutoff calibration factor} $\kappa_{\mathrm{RD}}^{\delta}(n,p,\nu,\alpha_0,c)$: out-of-sample validation performance (RMSE and MAE) of candidate representations across dimensions $p \in \{2, 5\}$ and degrees of freedom $\nu \in \{3, 5, 7\}$ ($\alpha_0 \in [0.10, 0.50]$, $n \in [50, 500]$).}
\label{tab:RD_interpolation_validation_full}
\footnotesize
\begin{tabular*}{\textwidth}{@{\extracolsep{\fill}} ccl cc @{}}
\toprule
$p$ & $\nu$ & \textbf{Candidate Method} & \textbf{Validation RMSE} & \textbf{Validation MAE} \\
\midrule
$2$ & $3$ & Linear Interpolation & $0.01695$ & $0.01172$ \\
    &     & Shape-preserving PCHIP & $0.01777$ & $0.01219$ \\
    &     & LOESS & $\mathbf{0.01185}$ & $\mathbf{0.00866}$ \\
\addlinespace[0.3em]
$2$ & $5$ & Linear Interpolation & $0.01400$ & $0.00932$ \\
    &     & Shape-preserving PCHIP & $0.01516$ & $0.00988$ \\
    &     & LOESS & $\mathbf{0.00855}$ & $\mathbf{0.00608}$ \\
\addlinespace[0.3em]
$2$ & $7$ & Linear Interpolation & $0.01186$ & $0.00771$ \\
    &     & Shape-preserving PCHIP & $0.01307$ & $0.00849$ \\
    &     & LOESS & $\mathbf{0.00815}$ & $\mathbf{0.00598}$ \\
\midrule
$5$ & $3$ & Linear Interpolation & $0.02838$ & $0.01403$ \\
    &     & Shape-preserving PCHIP & $0.02994$ & $0.01490$ \\
    &     & LOESS & $\mathbf{0.01454}$ & $\mathbf{0.00960}$ \\
\addlinespace[0.3em]
$5$ & $5$ & Linear Interpolation & $0.02074$ & $0.00985$ \\
    &     & Shape-preserving PCHIP & $0.02194$ & $0.01016$ \\
    &     & LOESS & $\mathbf{0.01123}$ & $\mathbf{0.00726}$ \\
\addlinespace[0.3em]
$5$ & $7$ & Linear Interpolation & $0.01944$ & $0.00840$ \\
    &     & Shape-preserving PCHIP & $0.02075$ & $0.00915$ \\
    &     & LOESS & $\mathbf{0.01124}$ & $\mathbf{0.00606}$ \\
\bottomrule
\end{tabular*}
\end{table}

\end{appendices}

\end{document}